\documentclass[reprint,
 amsmath,amssymb,
 aps,
prb,
]{revtex4-2}

\DeclareMathAlphabet{\mymathbb}{U}{bbold}{m}{n}

\usepackage[utf8]{inputenc}
\usepackage{mathtools,csquotes}
\usepackage{amssymb}
\usepackage{graphicx}
\usepackage{xcolor}
\usepackage{ulem}
\usepackage{bm}
\usepackage{hyperref}

\usepackage[makeroom]{cancel}

\renewcommand{\selectlanguage}[1]{}

\newcommand{\al}{\alpha}

\newcommand{\dd}{\mathrm{d}}
\newcommand{\ii}{\mathrm{i}}

\begin{document}

\preprint{APS/123-QED}

\title{Embedding 1D BDI topological dynamics into continuous elastic plates}

\author{Mohit Kumar}
\affiliation{Ray W. Herrick Laboratories, School of Mechanical Engineering, Purdue University, West Lafayette, Indiana 47907, USA}
\author{Fabio Semperlotti}%
\email{fsemperl@purdue.edu}
\affiliation{Ray W. Herrick Laboratories, School of Mechanical Engineering, Purdue University, West Lafayette, Indiana 47907, USA}

\date{\today}

\begin{abstract}
This study presents an approach that leverages the existing knowledge acquired in one-dimensional BDI class discrete metamaterials, such as mass-spring systems or acoustic resonators, and exploits it to realize fully continuous elastic two-dimensional topological waveguides. The design relies on the concept of evanescently coupled waveguides and defect resonances in order to reproduce the equivalent dynamics of prototypical BDI discrete systems, such as the Su-Schrieffer-Heeger (SSH) model. Starting with a continuous plate waveguide based on a periodic distribution of pillars, local resonators and waveguides are created by eliminating selected pillars and by exploiting the concept of point and line defects. The height of selected pillars is adjusted to tune the coupling strength between different resonators. The approach is validated by designing fully continuous elastic analogs of both the SSH chain and ladder systems. Numerical simulations and experimental results confirm the validity of the design by showing the emergence of topological edge modes at the interface of topologically distinct systems. In addition, the edge modes obtained in the elastic analog of the SSH ladder are shown to be Majorana-like modes.
\end{abstract}

\maketitle





\section{Introduction}
\label{sec:intro}
The general concept of topological band theory applied to classical metamaterials provides a range of methods to manipulate the propagation of waves in mechanical, acoustic, and elastic systems~\cite{huang_recent_2021,susstrunk_classification_2016,zhang_second_2023,yang_non-abelian_2024,shah_colloquium_2024}. This approach provides rational strategies to design localized topological edge modes robust against symmetry-preserving defects and imperfections, and it opens interesting practical applications such as energy harvesting, remote sensing, and vibration control~\cite{jiao_mechanical_2023,shah_colloquium_2024}. 

Topological edge modes are localized at interfaces between topologically distinct materials, which can be designed in a variety of methods~\cite{barlas_topological_2018,susstrunk_classification_2016,huang_recent_2021}. One possible approach, which is considered in the present work, is to emulate canonical topologically nontrivial systems corresponding to a specific class of the periodic table of topological insulators~(Tab.~\ref{tab:periodic-table-topological})~\cite{chiu_classification_2016,hasan_colloquium_2010,susstrunk_classification_2016}. Each class exhibits a unique physical behavior that manifests in the appearance of edge modes~\cite{chiu_classification_2016}. Examples include chiral Majorana modes in the 1D Su-Schrieffer-Heeger (SSH) model~\cite{su_soliton_1980} (BDI class), chiral Dirac modes in the 2D quantum Hall effect~\cite{klitzing_new_1980} (A class), helical Dirac modes 2D quantum spin Hall effect~\cite{kane_quantum_2005} (AII class), as well as their corresponding classical emulations~\cite{huang_recent_2021,shah_colloquium_2024}.
 
\begin{table}
    \caption{The periodic table of topological insulators for 1D, 2D, and 3D systems~\cite{chiu_classification_2016}. TRS: time-reveral symmetry, PHS: particle-hole symmetry, CS: chiral symmetry.}
    \begin{ruledtabular}
        \begin{tabular}{lcccccc}
            Class & \multicolumn{3}{c}{Symmetry} & \multicolumn{3}{c}{Invariants}\\
             & TRS & PHS & CS & 1D & 2D & 3D \\
            \hline A & 0 & 0 & 0 & 0 & $\mathbb{Z}$ & 0 \\
            AIII & 0 & 0 & 1 & $\mathbb{Z}$ & 0 & $\mathbb{Z}$ \\
            AI & $1$ & 0 & 0 & 0 & 0 & 0 \\
            BDI & $1$ & $1$ & 1 & $\mathbb{Z}$ & 0 & 0 \\
            D & 0 & $1$ & 0 & $\mathbb{Z}_2$ & $\mathbb{Z}$ & 0 \\
            DIII & $-1$ & $1$ & 1 & $\mathbb{Z}_2$ & $\mathbb{Z}_2$ & $\mathbb{Z}$ \\
            AII & $-1$ & 0 & 0 & 0 & $\mathbb{Z}_2$ & $\mathbb{Z}_2$ \\
            CII & $-1$ & $-1$ & 1 & $2 \mathbb{Z}$ & 0 & $\mathbb{Z}_2$ \\
            C & 0 & $-1$ & 0 & 0 & $2 \mathbb{Z}$ & 0 \\
            CI & $1$ & $-1$ & 1 & 0 & 0 & $2 \mathbb{Z}$ \\
        \end{tabular}
    \end{ruledtabular}
    \label{tab:periodic-table-topological}
\end{table}

Focusing on passive elastic systems, we note that they naturally possess time-reversal symmetry of the $+1$ type~\cite{susstrunk_classification_2016}. Indeed, the time-reversal operation $T$, defined as $t \to -t$ where $t$ is time, leaves the equations of linear elasticity invariant. Further, reversing time twice is nothing but the identity operation, that is, $T^2=1$.

Thus, in the present approach, passive elastic topological metamaterials are restricted to the AI, BDI, or CI classes. For nontrivial topological metamaterials to exist in a given spatial dimension, the corresponding entry in Tab.~\ref{tab:periodic-table-topological} must be nonzero. This restricts the choices of passive elastic topological metamaterials to 1D systems of the BDI class or 3D systems of the CI class.

Canonical 3D lattice models of the CI class require complicated imaginary~\cite{liu_elementary_2023} or nonlocal~\cite{schnyder_lattice_2009}
couplings. On the other hand, several 1D lattice models of the BDI class require simple nearest-neighbor couplings. Examples include the SSH model~\cite{su_soliton_1980} and the Kitaev chain model~\cite{kitaev_fault-tolerant_2003} (with real superconducting order parameter). From an engineering perspective, it is easier and practically more relevant to emulate structures of the latter type. Thus, in this work, we focus on 1D BDI class systems.

BDI class systems require classical analogs of particle-hole and chiral symmetries, which can be implemented by precisely controlling the coupling between different units of the system. These systems have been successfully realized using, for example, coupled acoustic cavities~\cite{chen_chiral_2020,yang_acoustic_2016,li_su-schrieffer-heeger_2018,guan_classical_2023} or mass-spring systems~\cite{hladky-hennion_experimental_2007,hladky-hennion_localized_2005,chaunsali_demonstrating_2017,allein_strain_2023,qian_observation_2023,prodan_dynamical_2017}.
It is certainly more challenging to control the coupling and implement particle-hole and chiral symmetries in continuous mechanical systems~\cite{chen_chiral_2020,chen_chiral_2024,coutant_acoustic_2021}. To the best of the authors' knowledge, there are currently no continuous elastic implementations of topological systems in the BDI class. The only study discussing a continuous BDI class system emulates the SSH model using an acoustic waveguide with segments of alternating heights~\cite{coutant_acoustic_2021}. 

Other topologically nontrivial continuous 1D systems in acoustics~\cite{xiao_geometric_2015,meng_designing_2018,yang_acoustic_2016,li_su-schrieffer-heeger_2018,zangeneh-nejad_topological_2019} and elasticity~\cite{yin_band_2018,coutant_surface_2024,kim_topologically_2017,zhang_zone_2019,wang_robust_2020,chaplain_topological_2020,huang_recent_2021}) rely on inversion symmetry and the resulting Zak phase invariant. In this case, interfaces between topologically distinct metamaterials can only support one edge mode whose frequency lies anywhere within the topological bandgap. This edge mode does not respect chiral or particle-hole symmetries. In contrast, as later shown, interfaces between topologically distinct metamaterials of the BDI class can support multiple edge modes with frequencies fixed at the center of the bandgap~\cite{chiu_classification_2016,prodan_dynamical_2017}. Further, such edge modes respect the chiral and particle-hole symmetries.

This study develops a general design strategy to embed 1D BDI class systems with nearest-neighbor couplings into 2D elastic waveguides. The basic design principle leverages evanescently coupled local resonances ~\cite{escalante_dispersion_2013,laude_principles_2021,jin_physics_2021,guo1DPhotonicTopological2024}, and it is analogous to the tight-binding method prevalent in quantum mechanics~\cite{simon_oxford_2013}. Local resonances are created inside the bandgap of an elastic plate featuring pillars arranged in a square lattice configuration; the coupling between resonances is tuned by adjusting the pillars' heights. This approach qualitatively reproduces chiral and particle-hole symmetries in a continuous elastic system. The validity and performance of this design method are illustrated by emulating the SSH model and an SSH ladder model.

The remainder of the paper is organized as follows. Section~\ref{sec:bdi-discrete-models} introduces general BDI class systems, the SSH model, and the dual SSH model. Section~\ref{sec:design-principle} presents our design principle. Sections~\ref{sec:elastic-ssh} and~\ref{sec:elastic-kitaev} apply the design principle to create elastic analogs of the SSH model and the dual SSH model. Section~\ref{sec:experiments} experimentally verifies the topological properties of the dual SSH model. Finally, Sec.~\ref{sec:conclusions} provides concluding remarks.

\section{Overview of discrete mechanical topological metamaterials in the BDI class}
\label{sec:bdi-discrete-models}
A typical BDI class system considered in this paper consists of $2N$ identical resonators arranged on a 1D bipartite lattice with nearest-neighbor couplings, where $N$ may be infinite. The bipartite nature of the lattice implies that the resonators can be divided into two sublattices (A and B) with $N$ resonators each, such that all couplings are between resonators of different sublattices. The state vector $\mathbf{U}$ of the system consists of the resonator displacements of sublattice A followed by resonator displacements of sublattice B, that is, $\mathbf{U}=(U_{1,A},\dots,U_{N,A}, U_{1,B},\dots,U_{N,B})^T$. The system is governed by the dynamical matrix $\mathbb{D}$ of size $2N \times 2N$, which can be written as
\begin{equation}
\label{eqn:hamiltonian}
    \mathbb{D} = \alpha \mymathbb{1}_{2N} \underbrace{-\begin{pmatrix}
        \mymathbb{0}_{N} & \mathbb{A}_{N \times N} \\
        \mathbb{A}_{N \times N}^{T} & \mymathbb{0}_{N}
    \end{pmatrix}\;}_{\widehat{\mathbb{D}}}
    ,
\end{equation}
where $\mymathbb{1}_{2N}$ is the identity matrix of size $2N$, $\mymathbb{0}_N$ is the zero matrix of size $N$, the superscript ${}^T$ denotes matrix transpose, $\widehat{\mathbb{D}}$ is the traceless part of the dynamical matrix, $\al$ is the natural frequency of the resonators, and $\mathbb{A}$ is a matrix of size $N \times N$ containing the details of the resonator interactions. The generic element $\mathbb{A}_{ij}$ equals the coupling strength between resonator $i$ of sublattice A and resonator $j$ of sublattice B. The negative sign in Eq.~\eqref{eqn:hamiltonian} ensures that $\mathbb{D}$ is analogous to the stiffness matrix of a spring-mass system. Although the coupling strengths can be any real number, we restrict all coupling strengths to be positive, as it is expected in discrete passive elastic systems. The modes and frequencies $f$ of the system are solutions of the eigenvalue problem
\begin{equation}
\label{eqn:evp}
    \mathbb{D} \mathbf{U} = f \mathbf{U}\;.
\end{equation}

The symmetries and topological properties of the system depend on the traceless dynamical matrix $\widehat{\mathbb{D}} = \mathbb{D} - \alpha \mymathbb{1}_{2N}$, since the term proportional to the identity matrix merely shifts the eigenvalues. For the system to be in the BDI class, it must admit time-reversal, chiral, and particle-hole symmetries of the $+1$ type. The system respects time-reversal symmetry of the $+1$ type because $\widehat{\mathbb{D}}$ is real~\cite{susstrunk_classification_2016,chiu_classification_2016}. The system respects chiral symmetry of the $+1$ type owing to the bipartite coupling scheme~\cite{padavic_topological_2018}. Mathematically, it satisfies the relation $\mathbb{U}^T \widehat{\mathbb{D}} \mathbb{U} = - \widehat{\mathbb{D}}$ for a matrix $\mathbb{U}$ that is unitary and obeys $\mathbb{U}^2=\mymathbb{1}$~\cite{chiu_classification_2016}.  Here, 
\begin{equation}
\label{eqn:chiral-operator}
    \mathbb{U} = 
    \begin{pmatrix}
        -\mymathbb{1}_N & \mymathbb{0}_N \\
        \mymathbb{0}_N & \mymathbb{1}_N
    \end{pmatrix}\;
\end{equation}
is the chiral operator that acts identically on a given sublattice. It reverses the displacement of resonators in sublattice A and leaves invariant the displacement of resonators in sublattice B. 

If a system respects time-reversal and chiral symmetries, it consequently respects particle-hole symmetry of the $+1$ type~\cite{prodan_dynamical_2017,susstrunk_classification_2016}, which means that $\widehat{\mathbb{D}}$ satisfies $\mathbb{V}^T \widehat{\mathbb{D}}^T \mathbb{V} = - \widehat{\mathbb{D}}$ for a matrix $\mathbb{V}$ that is unitary and obeys $\mathbb{V}^*\mathbb{V}=\mymathbb{1}$~\cite{chiu_classification_2016}, where the superscript $^*$ indicates the complex conjugate operation. $\mathbb{V}$ is called the particle-hole operator. Since $\widehat{\mathbb{D}}$ is real, the particle-hole operator equals the chiral operator~\cite{susstrunk_classification_2016}, that is,
\begin{equation}
\label{eqn:ph-operator}
    \mathbb{V}=\mathbb{U}=
    \begin{pmatrix}
        -\mymathbb{1}_N & \mymathbb{0}_N \\
        \mymathbb{0}_N & \mymathbb{1}_N
    \end{pmatrix}\;.
\end{equation}

Next, consider an infinite lattice where each unit cell has $2M$ resonators, with $M$ resonators in each sublattice. The modes of such a system are found by using the Floquet-Bloch ansatz~\cite{hussein_dynamics_2014}, according to which the modal displacements $\mathbf{U}_n$ in an arbitrary cell $n$ satisfy the relation $\mathbf{U}_n = \mathbf{u} e^{\ii k n}$, where $\mathbf{u}$ is the modal displacement in a reference unit cell (for which $n=0$), $k$ is the wavenumber, and $n$ is the number of unit cells from the reference. Substituting this ansatz in Eq.~\eqref{eqn:evp} we obtain a wavenumber-dependent eigenvalue problem 
\begin{equation}
\label{eqn:bloch-evp}
    \mathcal{D}(k) \mathbf{u} = f(k) \mathbf{u}\;, 
\end{equation}
where $\mathcal{D}(k)$ is the Bloch dynamical matrix of size $2M \times 2M$ and $f(k)$ is the dispersion relation. Explicitly,
\begin{equation}
\label{eqn:bloch-dynamical-matrix}
    \mathcal{D}(k) = \alpha \mymathbb{1}_{2M} - 
    \begin{pmatrix}
        \mymathbb{0}_M & \mathcal{A}(k)_{M \times M} \\
        \mathcal{A}(k)_{M \times M}^\dagger & \mymathbb{0}_M
    \end{pmatrix}\;,
\end{equation}
where
\begin{equation}
\label{eqn:bloch-coupling-matrix}
    \mathcal{A}_{ij}=\sum_{m=-\infty}^{\infty} \mathbb{A}_{i,j+mM}\:e^{ikm}\;
\end{equation}
is an $M \times M$ is the Bloch coupling matrix describing the resonator interactions and the superscript ${}^\dagger$ denotes the conjugate transpose operation.

The topological nature of the system is described by the winding number $\nu$, defined as~\cite{wakatsuki_fermion_2014} 
\begin{equation}
\label{eqn:winding-number}
    \nu = \frac{-1}{2 \pi \ii}\int_{-\pi}^{\pi} \partial_k \ln{\mathrm{Det}(\mathcal{A}(k))} \, \dd k\;.  
\end{equation}
The winding number measures the number of clockwise loops the determinant of $\mathcal{A}(k)$ traces around the origin of the complex plane. It is generally unaffected by small symmetry-preserving perturbations, such as small variations in the strength of the resonator interactions. In general, symmetry-preserving perturbations are those that maintain the time-reversal symmetry and the bipartite coupling scheme of the system. This robustness follows from the topological nature of the winding number.

The winding number indicates the appearance of topological edge modes via the bulk-boundary correspondence principle~\cite{chiu_classification_2016}. These edge modes arise at the interface of two semi-infinite BDI systems with identical values of $\al$ or at the free end of a semi-infinite BDI system. At an interface, the number of topological edge modes  $N_{\mathrm{edge}}$ equals the difference between the winding numbers $\nu_1$ and $\nu_2$ of the two constituent BDI systems. At a free end, $\nu_2=0$ and the number of topological edge modes $N_{\mathrm{edge}}$ equals the winding number $\nu_1$ of the BDI system. In summary,
\begin{equation}
    N_{\mathrm{edge}} = |\nu_1 - \nu_2|\;.
\end{equation}

The frequency and displacement characteristics of the topological edge modes are dictated by the symmetries of the BDI class. The frequency of an edge mode is fixed at $\al$ and its displacement is confined to one of the two sublattices of resonators~\cite{chiu_classification_2016}. Thus, using Eqs.~\eqref{eqn:chiral-operator},~\eqref{eqn:ph-operator}, it follows that a topological edge mode $\mathbf{U}_\mathrm{edge}$ is invariant up to a sign factor under the chiral and particle-hole operators, that is, $\mathbb{U} \mathbf{U}_\mathrm{edge} = \pm \mathbf{U}_\mathrm{edge}$ and  $\mathbb{V} \mathbf{U}_\mathrm{edge} = \pm \mathbf{U}_\mathrm{edge}$.

In the following, we revisit two prototypical BDI class models and their topological properties. These two systems will form the foundation of the continuous elastic designs presented in Sec.~\ref{sec:elastic-ssh} and Sec.~\ref{sec:elastic-kitaev}.

\subsection{Su-Schrieffer-Heeger model}
\label{sec:ssh}
\begin{figure}
    \centering
    \includegraphics[width=\linewidth]{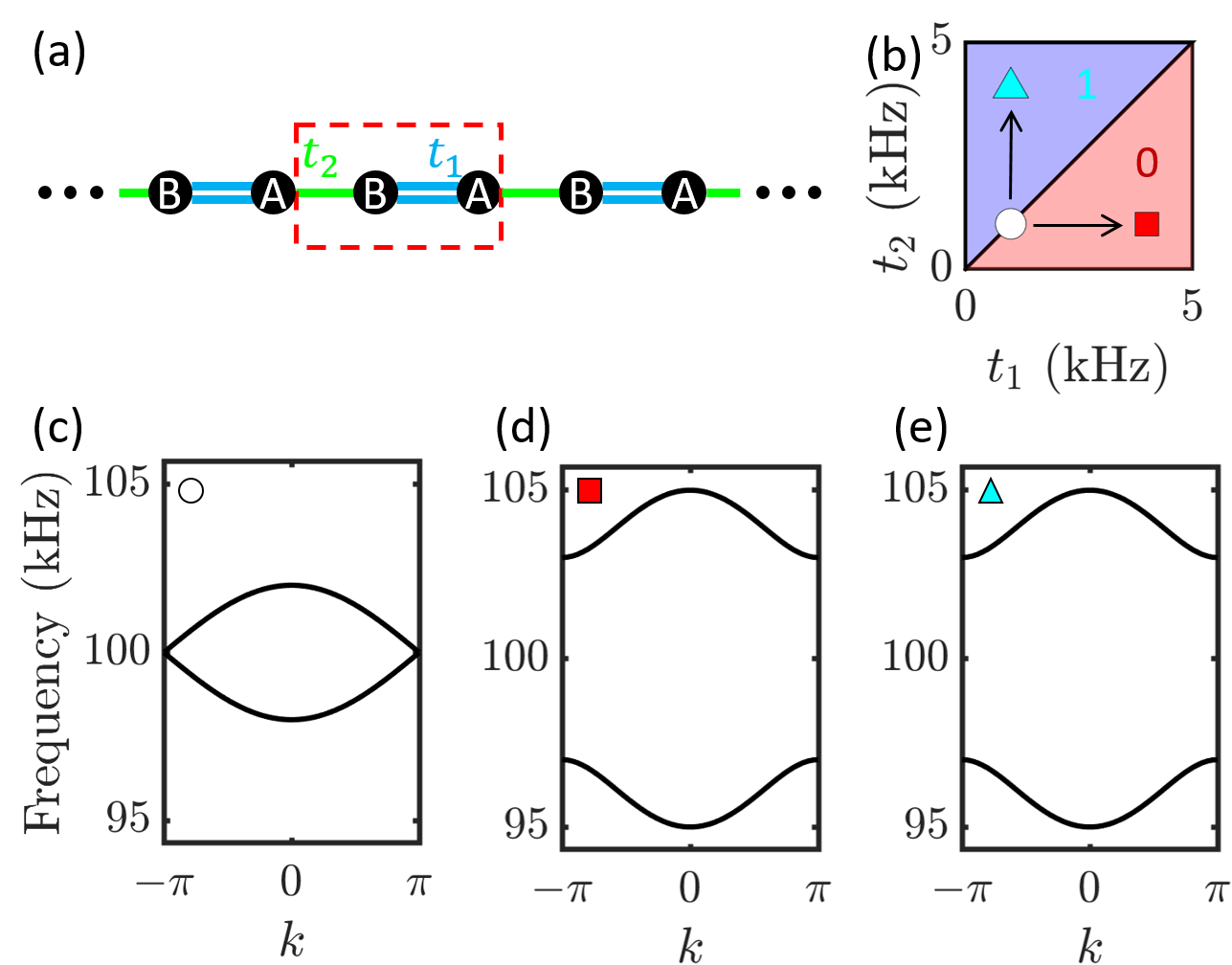}
    \caption{(a) The SSH model realized using identical resonators. The dashed rectangle marks one unit cell. The letters mark sublattices A and B. (b) Topological phase diagram of the SSH model expressed in terms of the coupling strengths. The two regions are marked by their winding numbers, $0$ and $1$. The dispersion relations corresponding to the circle, square, and triangle markers are shown in (c-e). ($\al$ is fixed at $100$ kHz.)}
    \label{fig:ssh}
\end{figure}

The Su-Schrieffer-Heeger (SSH) model~\cite{su_soliton_1980} consists of a repeating arrangement of two identical resonators with alternating coupling strengths $t_1$ and $t_2$ (Fig.~\ref{fig:ssh}a). The coupling matrix for a finite system of $2N$ resonators is \begin{equation}
    \mathbb{A}_{N \times N} = 
    \begin{pmatrix}
        t_1 & 0 & \cdots & \cdots & 0 \\
        t_2 & t_1 & 0 & \cdots & 0\\
        \cdots & \cdots & \cdots & \cdots & \cdots \\
        0 & \cdots & 0 & t_1 & 0\\
        0 & \cdots & \cdots & t_2 & t_1
    \end{pmatrix}\;.
\end{equation}

\subsubsection{Topological properties of the Bloch dynamical matrix}

For a periodic lattice, the Floquet-Bloch ansatz results in the following Bloch dynamical matrix:
\begin{equation}
\label{eqn:ssh-dynamical-matrix}
    \mathcal{D}(k)= \alpha \mymathbb{1} -
    \begin{pmatrix}
        0 & t_1 + t_2 e^{-\ii k} \\
        t_1 + t_2 e^{\ii k} & 0
    \end{pmatrix}\;,
\end{equation}
where $\mathcal{A}(k)=t_1 + t_2 e^{-\ii k}$ from Eq.~\eqref{eqn:bloch-coupling-matrix}. The dispersion curves are obtained by solving Eq.~\eqref{eqn:bloch-evp} as
\begin{equation}
\label{eqn:ssh-dispersion-relation}
    f_{\pm}(k) = \al \pm \sqrt{t_1 ^2 + t_2^2 + 2 t_1 t_2 \cos k}\;.
\end{equation}

The relation between the dispersion curves and the $(t_1,t_2)$ parameter space is illustrated in Figs.~\ref{fig:ssh}b-e. If $t_1=t_2$, the dispersion curves are degenerate at $k=\pi$. For example, the parameters $\al=100$ kHz, $t_1=1$ kHz, and $t_2=1$ kHz (white circle in Fig.~\ref{fig:ssh}b) result in the dispersion curves in Fig.~\ref{fig:ssh}c. If $t_1 \neq t_2$, the dispersion curves are separated by a bandgap centered at $\al$. For example, the parameters $\al=100$ kHz, $t_1=4$ kHz, and $t_2=1$ kHz (red square) result in the dispersion curves in Fig.~\ref{fig:ssh}d, and the parameters $\al=100$ kHz, $t_1=4$ kHz, and $t_2=1$ kHz (blue triangle) result in the dispersion curves in Fig.~\ref{fig:ssh}e. Thus, for a given $\al$, the parameter space is split into two regions by the line of degenerate parameter values, as shown in Fig.~\ref{fig:ssh}b. 

Parameter sets from different regions create SSH models with different winding numbers. To compute the winding number of the system, note that $\mathrm{Det}\mathcal{A}(k)=\mathcal{A}(k)=t_1 + t_2 e^{-\ii k}$, which traces out a circle of radius $t_2$ centered about $t_1$ in a clockwise manner. Thus, by applying Eq.~\eqref{eqn:winding-number},
\begin{equation}
\label{eqn:ssh-winding-number}
    \nu = 
    \begin{cases}
        0,& t_1>t_2\\
        1,& t_1<t_2\;.
    \end{cases}
\end{equation}

\subsubsection{Topological edge modes}
\begin{figure}
    \centering
    \includegraphics[width=\linewidth]{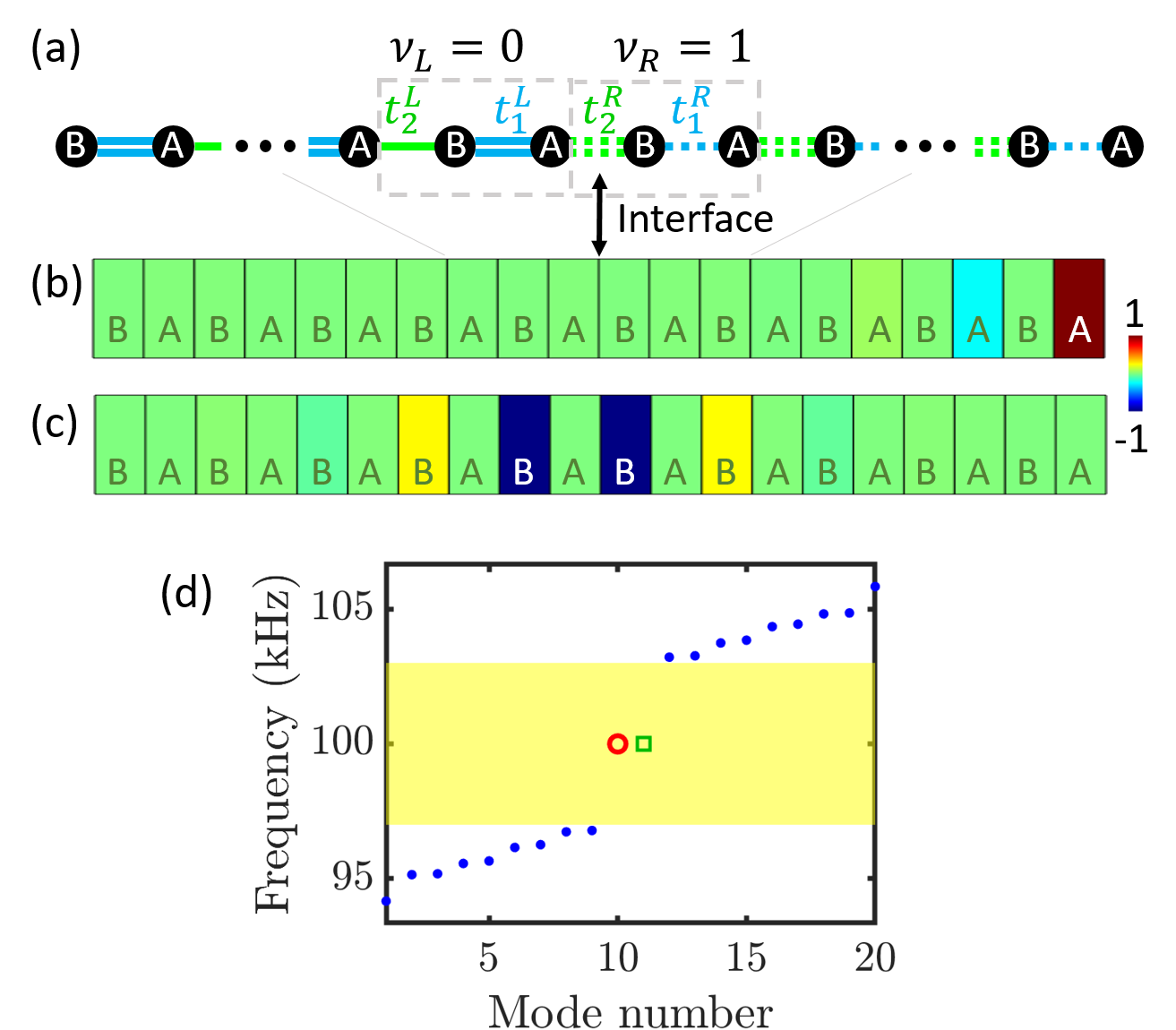}
    \caption{(a) Interface between two finite SSH chains with winding numbers $0$ and $1$. The letters mark the two sublattices. (b,c) Topological edge modes localized at the right end and at the interface. Each rectangle denotes a resonator, whose color indicates its displacement and whose letter marks the sublattice. The displacements are scaled so that the largest magnitude is 1. (d) Natural frequencies of the system. The red circle, green square, and blue dots represent a topological edge mode at the right end, a topological edge mode at the interface, and bulk modes. The yellow rectangle marks the common bandgap of the left and right SSH chains.}
    \label{fig:ssh-edge-mode}
\end{figure}

The winding number of the SSH model leads to the existence of edge modes in a finite system because of the bulk-boundary correspondence~\cite{chiu_classification_2016}. Topological edge modes appear at the free end of a finite SSH chain with winding number 1 or at the interface of two SSH chains with different winding numbers. To investigate these edge modes, consider a system obtained by joining two finite SSH chains of 10 unit cells each, denoted left (L) and right (R), as shown in Fig.~\ref{fig:ssh-edge-mode}a. Both chains have identical values of $\al$. In the left chain, $t^L_1 > t^L_2$, leading to a winding number $\nu_L = 0$, while in the right chain, $t^R_1 < t^R_2$, leading to a winding number $\nu_R=1$. We choose $\al=100$ kHz, $t^L_1=4$ kHz, $t^L_2=1$ kHz, $t^R_1=1$ kHz, and $t^R_2=4$ kHz. The natural frequencies of the system are shown in Fig.~\ref{fig:ssh-edge-mode}d. There are two eigenmodes with frequency $\al=100$ kHz (red circle and green square) that are separated from the bulk modes (blue dots). These eigenmodes are the topological edge modes, which are plotted in Fig.~\ref{fig:ssh-edge-mode}b and Fig.~\ref{fig:ssh-edge-mode}c. One of the topological edge modes (red circle) is confined to the right end, while the other mode (green square) is confined to the interface. There is no topological mode confined to the left end. These results are in agreement with the bulk-boundary correspondence. 

The displacements of edge modes exhibit symmetries characteristic of the BDI class. First, consider the edge mode at the right end $\mathbf{U}_\mathrm{right}$, which is shown in Fig.~\ref{fig:ssh-edge-mode}b. The displacements are nonzero for resonators in sublattice A and zero for resonators in sublattice B. As a consequence, when the chiral operator (Eq.~\eqref{eqn:chiral-operator}), which reverses the displacements of sublattice A, acts on the mode shape, it merely changes the sign of the mode shape. That is, $\mathbb{U}\mathbf{U}_\mathrm{right}=-\mathbf{U}_\mathrm{right}$. Thus, the mode shape is invariant up to a sign factor. Similarly, the particle-hole operator (Eq.~\eqref{eqn:ph-operator}) leaves the mode shape invariant up to a sign factor. 

Next, consider the edge mode at the interface $\mathbf{U}_\mathrm{int}$, which is plotted in Fig.~\ref{fig:ssh-edge-mode}c. The displacements are nonzero for resonators in sublattice B and zero for resonators in sublattice A. Thus, the chiral and particle-hole operators leave the mode shape invariant. That is, $\mathbb{U}\mathbf{U}_\mathrm{int}=\mathbf{U}_\mathrm{int}$ and $\mathbb{V}\mathbf{U}_\mathrm{int}=\mathbf{U}_\mathrm{int}$.

This discussion of the SSH model highlights several features imparted by chiral and particle-hole symmetries: (i) The dispersion relations of the infinite lattices (Figs.~\ref{fig:ssh}c-e) are symmetric about the center frequency $\al$~\cite{susstrunk_classification_2016}; (ii) The topological edge modes have a frequency equal to $\al$ (Fig.~\ref{fig:ssh-edge-mode}c); (iii) The displacements of the edge mode are confined to one sublattice of resonators  (Fig.~\ref{fig:ssh-edge-mode}b); (iv) The topological edge modes are invariant up to a sign factor under chiral and particle-hole operators. These criteria will be helpful when evaluating the role of chiral symmetry for continuous systems presented in later sections. 

\subsection{Dual SSH model}
\label{sec:dual-ssh}
\begin{figure}
    \centering
    \includegraphics[width=\linewidth]{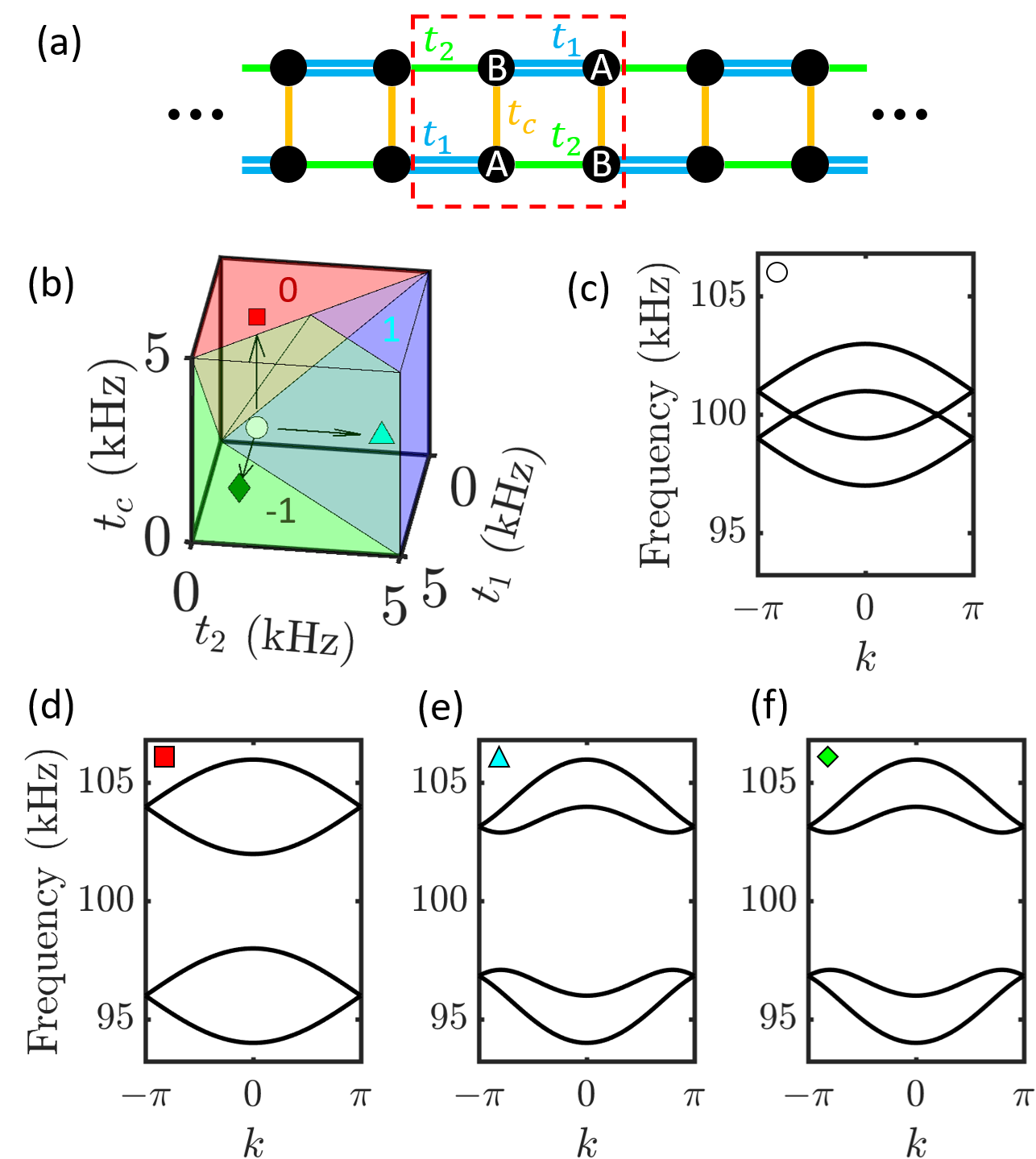}
    \caption{(a) The dual SSH model realized using identical resonators. The dashed rectangle marks one unit cell. The letters mark sublattices A and B. (b) Topological phase diagram of the dual SSH model. The three regions are marked by their winding numbers, $0$, $1$, and $-1$. The planes separating the regions consist of degenerate parameter values. The white circle lies on the plane $t_1=t_2$. Perturbing it along the $t_1$, $t_2$, or $t_c$ axes moves it into region 1 (blue triangle), region $-1$ (green diamond), or region $0$ (red square). The dispersion relations corresponding to each marker is plotted in (c)-(e). ($\al$ is fixed at $100$ kHz.)}
    \label{fig:dssh-plots}
\end{figure}

The dual SSH model is an SSH ladder consisting of two staggered SSH chains coupled to each other (Fig.~\ref{fig:dssh-plots}a)~\cite{padavic_topological_2018,qian_observation_2023,wakatsuki_fermion_2014,li_extended_2019}. The intra-SSH coupling strengths are $t_1$ and $t_2$, and the inter-SSH coupling strength is $t_c$. The coupling matrix for a finite dual SSH chain with $2N$ resonators is 
\begin{equation}
\label{eqn:dual-ssh-coupling-matrix}
    \mathbb{A}_{N \times N} =  
    \begin{pmatrix}
        t_c & t_1 & \cdots & \cdots & 0 \\
        t_2 & t_c & t_1 & \cdots & 0\\
        \cdots & \cdots & \cdots & \cdots & \cdots \\
        0 & \cdots & t_2 & t_c & t_1 \\ 
        0 & \cdots & \cdots & t_2 & t_c
    \end{pmatrix}\;.
\end{equation}

The dynamical matrix of the dual SSH model, under a change of basis by the matrix 
\begin{equation}
    \mathbb{W}=\frac{1}{\sqrt{2}}
    \begin{pmatrix}
        \mymathbb{1}_N & -\mymathbb{1}_N \\
        \mymathbb{1}_N & \mymathbb{1}_N
    \end{pmatrix}\;,
\end{equation}
is equivalent to the dynamical matrix of a classical analog of the Kitaev chain model with real superconducting order parameter~\cite{qian_observation_2023,padavic_topological_2018,guan_classical_2023,guo_observation_2021} (Supplementary Material Sec.~I~C). Thus, investigating the dual SSH model is equivalent to investigating a (restricted) Kitaev chain model. 

\subsubsection{Topological properties of the Bloch dynamical matrix}
For a periodic lattice of the dual SSH chain, the Floquet-Bloch ansatz leads to the Bloch dynamical matrix (Eq.~\eqref{eqn:bloch-dynamical-matrix}) with the Bloch coupling matrix
\begin{equation}
\label{eqn:bloch-coupling-matrix-dual-ssh}
    \mathcal{A}(k) = 
    \begin{pmatrix}
        t_c & t_1 + t_2 e^{-\ii k} \\
        t_2 + t_1 e^{\ii k} & t_c
    \end{pmatrix}
\end{equation}
from Eq.~\eqref{eqn:bloch-coupling-matrix}.

The wavenumber-dependent eigenvalue problem in Eq.~\eqref{eqn:bloch-evp} can be solved to obtain the dispersion relations, which read~\cite{padavic_topological_2018}
\begin{multline}
\label{eqn:dssh-dispersion-relation}
    f(k) = \al \pm \left( t_1^2 + t_2^2 + \mu^2 + 2t_1 t_2 \cos(k) \right.\\ \left. \pm 2(t_1+t_2)t_c \cos(k/2) \right)^{\frac{1}{2}}\;.
\end{multline}

Figure~\ref{fig:dssh-plots} illustrates the relation between the dispersion curves and the model parameters. Consider the three parameter sets marked in Fig.~\ref{fig:dssh-plots}b: $\al=100$ kHz, $t_1=1$ kHz, $t_2=1$ kHz, $t_c=4$ kHz (red square); $\al=100$ kHz, $t_1=1$ kHz, $t_2=4$ kHz, $t_c=1$ kHz (blue triangle); and $\al=100$ kHz, $t_1=4$ kHz, $t_2=1$ kHz, $t_c=1$ kHz (green diamond). The dispersion relations of the corresponding dual SSH models are shown in Figs.~\ref{fig:dssh-plots}d-f. In all plots, the two upper and two lower dispersion curves are separated by a bandgap centered around $\alpha$. In addition, the two upper and two lower dispersion curves are two-fold degenerate at $k=\pi$. This degeneracy is because of the zone-folding effect~\cite{zhang_zone_2019}, which results from choosing a unit cell with four resonators, which is larger than the smallest possible one with two resonators.

There are additional degeneracies for specific parameter values~\cite{li_extended_2019,padavic_topological_2018}. On the plane $t_1 = t_2$ and $t_c < t_1 + t_2$, the bandgap closes at $k=\pm \cos^{-1}(-t_c/(t_1 + t_2))$. For example, the white circle in Fig.~\ref{fig:dssh-plots}b with parameters $t_1=1$ kHz, $t_2=1$ kHz, and $t_c=1$ kHz results in the dispersion curves shown in Fig.~\ref{fig:dssh-plots}c. In addition, when $t_c=t_1 + t_2$, the bandgap closes at $k=0$. The planes of degenerate parameter values divide the parameter space into the three regions shown in Fig.~\ref{fig:dssh-plots}b. 

Dual SSH chains with parameter values from different regions are topologically distinct because they have different winding numbers. Indeed, $\mathrm{Det} \mathcal{A}(k) = t_c^2 - 2 t_1 t_2 - t_1^2 e^{\ii k} - t_2^2 e^{-\ii k}$ (from Eq.~\eqref{eqn:bloch-coupling-matrix-dual-ssh}) traces an ellipse in the complex plane with winding number
\begin{equation}
\label{eqn:dual-ssh-winding-number}
    \nu = 
    \begin{cases}
        0, & t_c > t_1 + t_2\\
        1, & t_c < t_1 + t_2 \text{ and } t_1 < t_2 \\
        -1, & t_c < t_1 + t_2 \text{ and } t_1 > t_2 \;.
    \end{cases}
\end{equation}

\subsubsection{Topological edge modes}
The winding numbers of the dual SSH chains manifest themselves as edge modes in a finite system~\cite{padavic_topological_2018}. According to the bulk-boundary correspondence~\cite{chiu_classification_2016}, there are five configurations that support topological edge modes: (i) the free end of a dual SSH chain with winding number $1$, (ii) the free end of a dual SSH chain with winding number $-1$, (iii) the interface between dual SSH chains with winding numbers $0$ and $1$, (iv) the interface between dual SSH chains with winding numbers $0$ and $-1$, and (v) the interface between dual SSH chains with winding numbers $1$ and $-1$. 

The various edge modes can be investigated in a finite system constructed by joining two finite chains of the dual SSH model, denoted left (L) and right (R), with 5 unit cells each. The two chains have different winding numbers, $\nu_L$ and $\nu_R$. We consider two cases: $\nu_L=0$, $\nu_R=1$ and $\nu_L=1$, $\nu_R=-1$. (A third case $\nu_L=0$, $\nu_R=-1$ is neglected because its edge modes are similar to $\nu_L=0$, $\nu_R=1$.) 

\begin{figure}
    \centering
    \includegraphics[width=\linewidth]{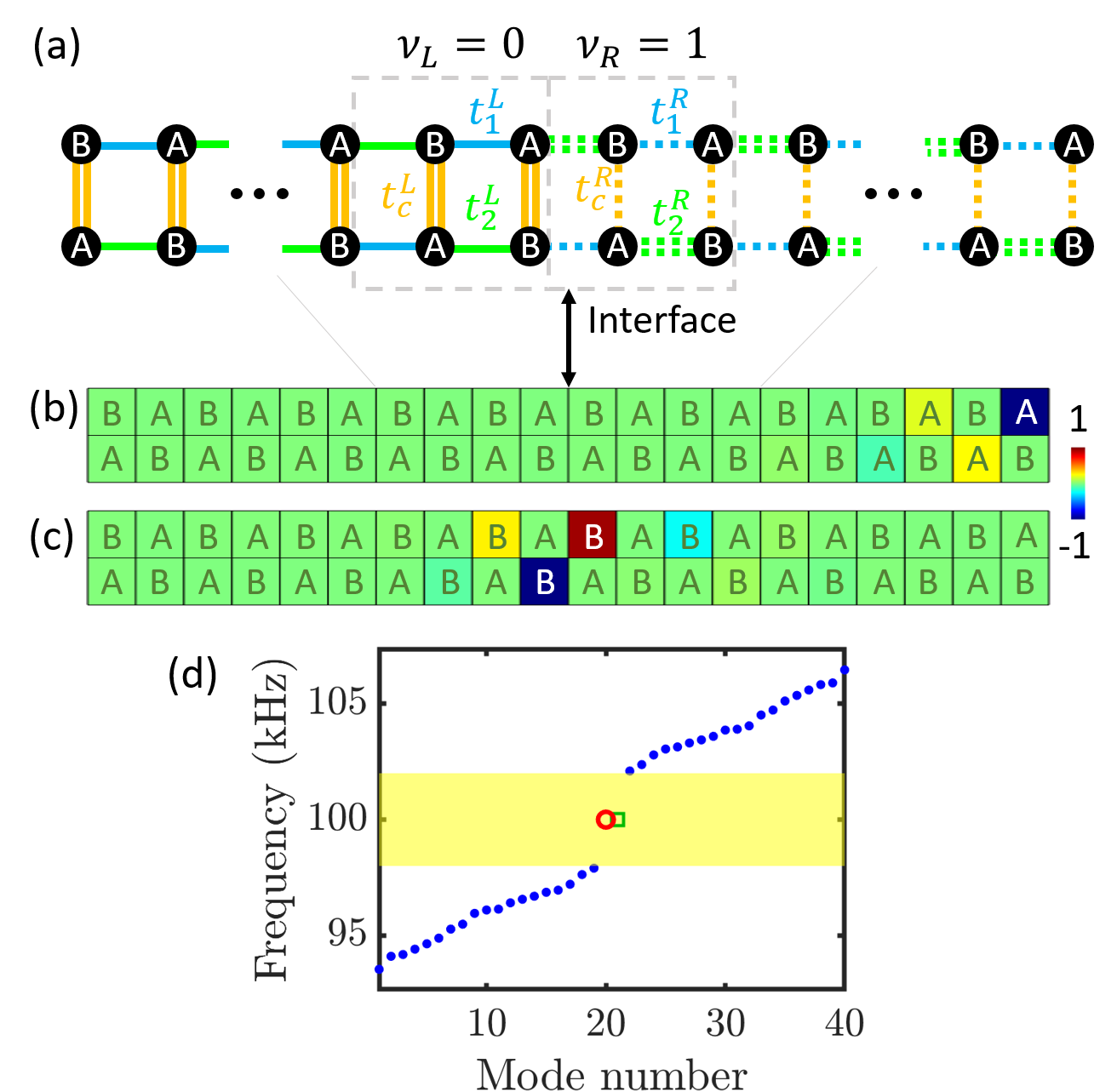}
    \caption{(a) The interface between two finite dual SSH models with winding numbers $0$ and $1$. (b,c) Topological edge modes localized at the right end and at the interface. Each rectangle denotes a resonator, whose color corresponds to its displacement and whose letter marks the sublattice. The displacements are scaled so that the largest magnitude is 1. (d) Natural frequencies of the system. The red circle, green square, and blue dots represent a topological edge mode at the interface, a topological edge mode at the right end, and bulk modes. The yellow rectangle marks the common bandgap of the left and right dual SSH models.}
    \label{fig:dssh-edge-mode-10}
\end{figure}

The first system is shown in Fig.~\ref{fig:dssh-edge-mode-10}a. For the numerical realization, the center frequency is fixed at $\al=100$ kHz. For the left chain, we choose $t^L_1=1$ kHz, $t^L_2=1$ kHz, and $t^L_c=4$ kHz, leading to a winding number $\nu_L = 0$. For the right chain, we choose $t^R_1=1$ kHz, $t^R_2=4$ kHz, and $t^R_c=1$ kHz, leading to a winding number $\nu_R = 1$. The natural frequencies of the system are shown in Fig.~\ref{fig:dssh-edge-mode-10}d. There are two topological edge modes (red circle and green square) with frequency $\al$ that are separated from the bulk modes (blue dots). One of the topological edge modes (red circle) is confined to the right end (Fig.~\ref{fig:dssh-edge-mode-10}b), while the other topological edge mode (green square) is confined to the interface (Fig.~\ref{fig:dssh-edge-mode-10}c). There is no topological mode confined to the left end. These results are in agreement with the bulk-boundary correspondence~\cite{chiu_classification_2016}.

The mode shapes of the topological edge modes exhibit the characteristic symmetries of the BDI class. For the topological edge mode localized at the right end (Fig.~\ref{fig:dssh-edge-mode-10}b), the displacements are nonzero only for resonators of sublattice A, due to which the mode shape merely changes sign under chiral and particle-hole operators. Similarly, for the topological edge mode localized at the interface (Fig.~\ref{fig:dssh-edge-mode-10}c), the displacements are nonzero only for resonators of sublattice B, due to which the mode shape is invariant under chiral and particle-hole operators.

Furthermore, recall that the dual SSH chain and the Kitaev chain model differ only by a change of basis by the matrix $\mathbb{W}$ (Supplementary Material Sec.~I~C). Upon applying the transformation $\mathbb{W}$ to the topological edge mode localized at the right end, it maps exactly to the Majorana-like mode supported at the free end of a classical analog of the topologically nontrivial Kitaev chain (Supplementary Material Sec.~I~D). Similarly, the topological edge mode localized at the interface corresponds to the Majorana-like mode supported at the interface between topologically trivial and nontrivial Kitaev chains (Supplementary Material Sec.~I~D). Thus, we refer to these topological edge modes of the dual SSH as Majorana-like modes.

We use the term Majorana-like modes for classical systems to emphasize fundamental differences when compared against Majorana fermions studied in condensed matter physics. In the latter systems, the constituent \enquote{resonators} are fermions, which satisfy Fermi-Dirac statistics and are described by creation and annihilation operators satisfying fermion anticommutation relations~\cite{altland_condensed_2010}. They lead to chiral Majorana fermions in BDI class systems and Majorana fermions in D class systems~\cite{chiu_classification_2016}, which are described by Majorana creation and annihilation operators obeying the Majorana anticommutation relations~\cite{sarma_majorana_2015}. However, there are no straightforward analogs of creation and annihilation operators and quantum statistics for classical systems~\cite{liu_classical_2023,prodan_dynamical_2017}.

\begin{figure}
    \centering
    \includegraphics[width=\linewidth]{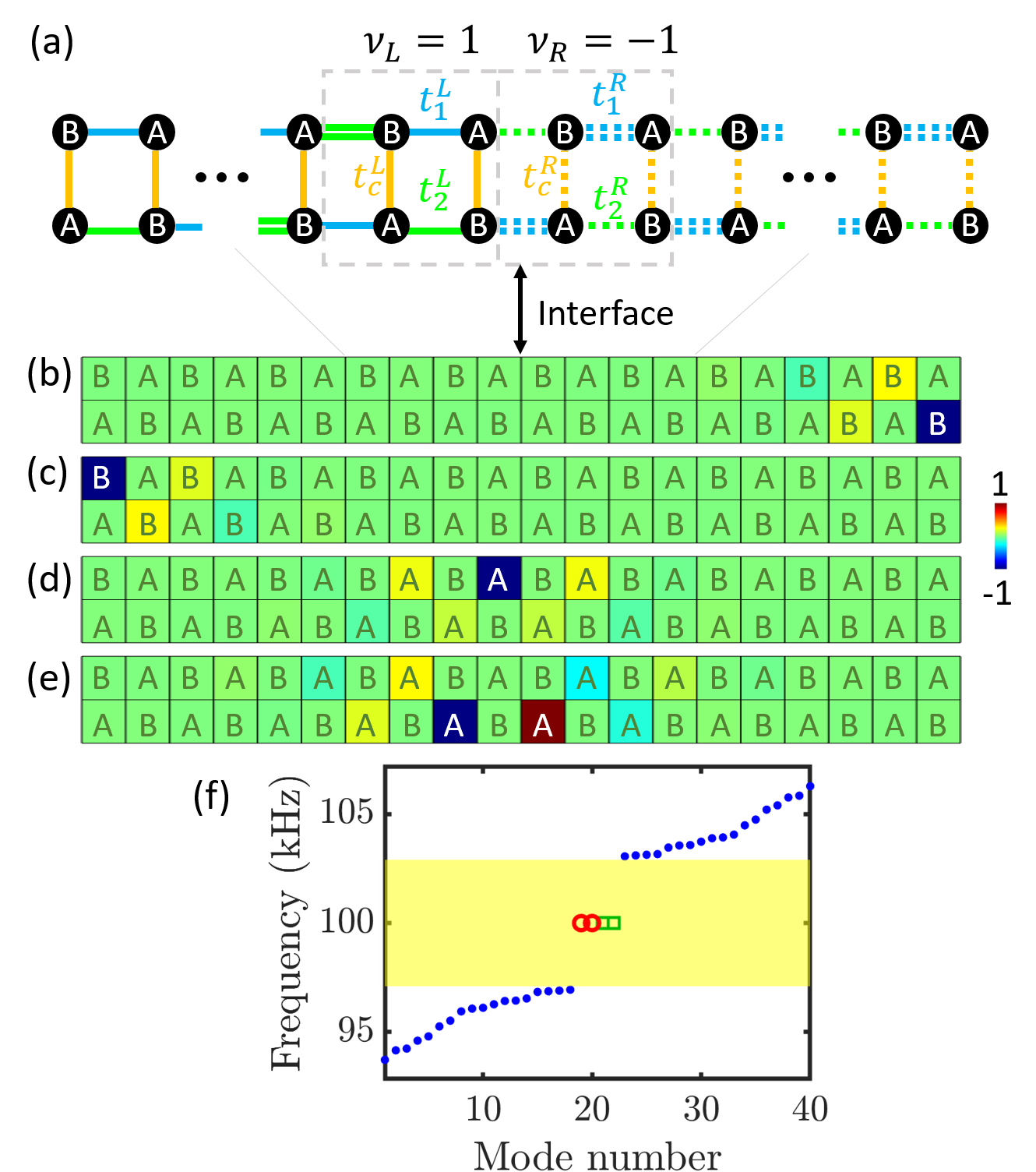}
    \caption{(a) The interface between two finite dual SSH models with winding numbers $1$ and $-1$. (b-e) Topological edge modes localized at the right end, left end, and interface. Each rectangle denotes a resonator, whose color corresponds to its displacement and whose letter marks the sublattice. The displacements are scaled so that the largest magnitude is 1. (f) Natural frequencies of the system. The red circles, green squares, and blue dots represent edge modes at the interface, edge modes at the ends, and bulk modes. The yellow rectangle marks the common bandgap of the left and right dual SSH models.}
    \label{fig:dssh-edge-modes-1n1}
\end{figure}

The second system is shown in Fig.~\ref{fig:dssh-edge-modes-1n1}a, which features an interface between dual SSH chains with winding numbers $\nu_L=1$ and $\nu_R=-1$. Its dynamics are simulated in the same manner by choosing the parameter values $\al=100$ kHz, $t^L_1=1$ kHz, $t^L_2=4$ kHz, $t^L_c=1$ kHz, $t^R_1=4$ kHz, $t^R_2=1$ kHz, and $t^R_c=1$ kHz. Figure~\ref{fig:dssh-edge-modes-1n1}f plots the natural frequencies, where four topological edge modes have frequency $\al$. One topological mode is localized at the right end (Fig.~\ref{fig:dssh-edge-modes-1n1}b), one topological mode is localized at the left end (Fig.~\ref{fig:dssh-edge-modes-1n1}c), and two topological modes are localized at the interface (Fig.~\ref{fig:dssh-edge-modes-1n1}d,e). These results agree with the bulk-boundary correspondence~\cite{chiu_classification_2016}. The edge modes localized on the right or left ends are confined to sublattice B, while the edge modes localized at the interface are confined to sublattice A. Thus, the modes are invariant up to a sign factor under chiral and particle-hole operations.

\section{Topological continuous waveguides: embedding discrete topological models in continuous systems}
\label{sec:design-principle}

Emulating BDI class models using continuous elastic systems poses several challenges including: (i) implementing a bipartite coupling scheme between waveguide modes that preserves chiral and particle-hole symmetries, (ii) controlling the interactions between waveguide modes, and (iii) choosing parameters to achieve desired coupling strengths. To overcome these challenges, we introduce a general design platform based on an engineered 2D waveguide and then embed discrete resonator models into it. 

\subsection{Design specifications of the 2D engineered waveguide}
\label{sec:design-specifications}
\begin{figure}
    \centering
    \includegraphics[width=\linewidth]{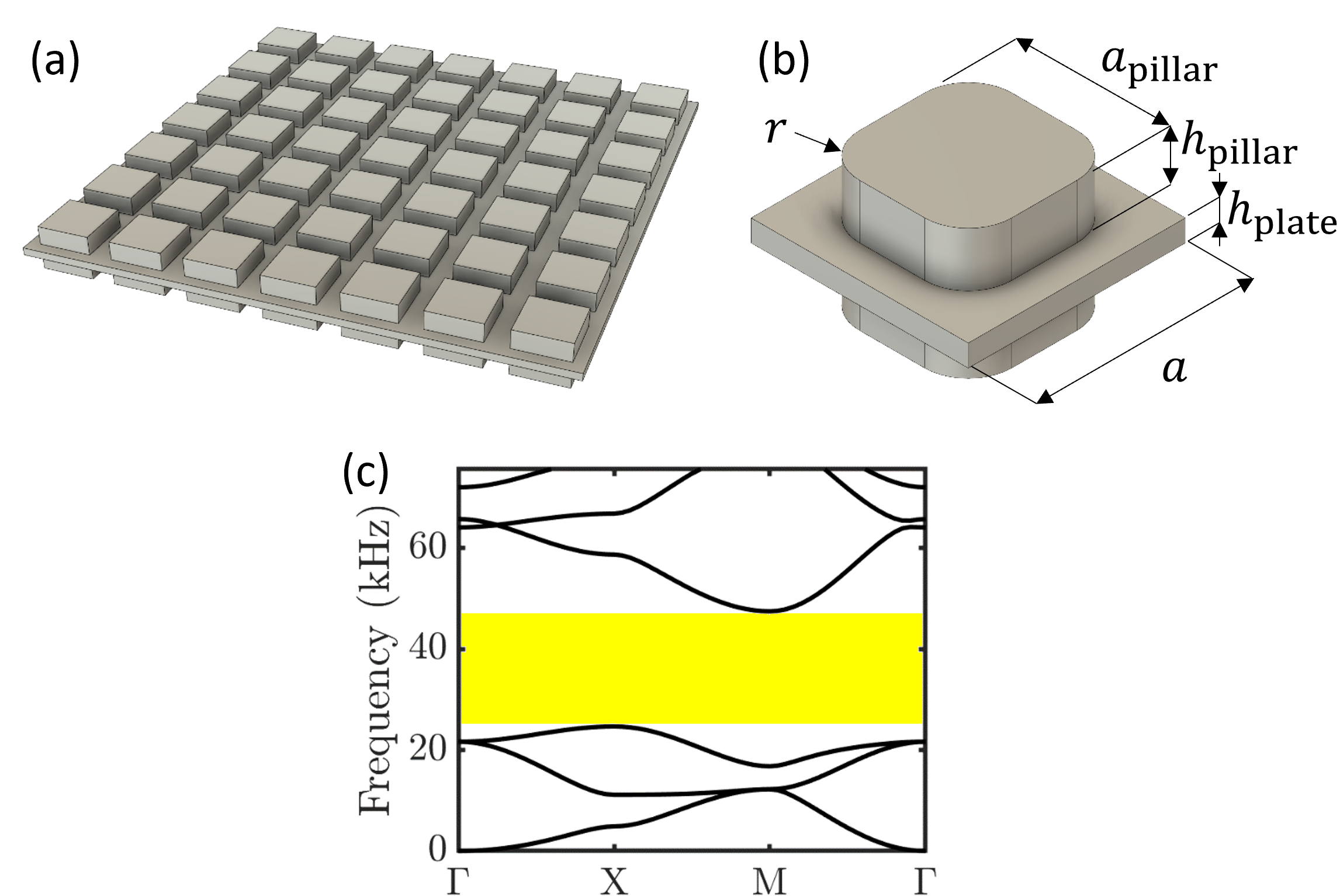}
    \caption{(a) An elastic metamaterial consisting of square pillars without fillets ($r=0$ in) arranged on a plate. (b) Isometric view of a general unit cell with fillets. (c) Dispersion relation of flexural modes of the metamaterial with $r=0$ in. The bandgap is highlighted in yellow.}
    \label{fig:bandgap-plate}
\end{figure}

The 2D engineered waveguide consists of a continuous thin plate with a periodic lattice of pillar structures built on both sides, as shown in Fig.~\ref{fig:bandgap-plate}a. The waveguide is symmetric about the midplane. More specifically, in the following we will consider a plate of thickness $h_{\rm{plate}}$ with square pillars arranged in a square grid with spacing $a$. The square pillars have a side length $a_{\rm{pillar}}$, height $h_{\rm{pillar}}$ (above the plate), and corner fillet radius $r$ (Fig.~\ref{fig:bandgap-plate}b). For the subsequent simulations, the geometric parameters will take the following values: $h_{\rm{plate}}=0.125$ in (3.175 mm), $h_{\rm{pillar}}=0.3125$ in (7.938 mm), $a=1.25$ in (31.75 mm), and $a_{\rm{pillar}}=0.875$ in (22.225 mm). The plate is made of aluminum with density $\rho=2700$ kg/m$^3$, Young's modulus of elasticity $E=70$ GPa, and Poisson's ratio as $1/3$. The parameter $r$ will be explicitly provided for specific analyses. 

This waveguide is a classical configuration that results in the formation of bandgaps~\cite{wu_evidence_2008,jin_physics_2021}. Figure~\ref{fig:bandgap-plate}c shows the dispersion relation and the bandgap assuming $r=0$ in. To improve clarity and also because they will be the main focus of this study, only the flexural modes (that is, the asymmetric Lamb modes) are shown. 

The dispersion relations are computed by finite element simulations of a unit cell via the commercial software COMSOL Multiphysics 6.1. Assume that the plate lies on the $xz$ plane and the unit cell occupies the planar region $[x_1,x_2] \times [z_1,z_2]$. The differential eigenvalue problem to be solved is defined by the governing equations of linear elasticity for an isotropic material,
\begin{equation}
\label{eqn:navier-lame}
    \mu \nabla^2 \mathbf{u} + (\lambda + \mu) \nabla (\nabla \cdot \mathbf{u})=-\rho \omega^2 \mathbf{u}\;\;,
\end{equation}
where $\mathbf{u}$ is the displacement field and $\lambda$ and $\mu$ are Lam\'e constants. Floquet-Bloch boundary conditions are applied on the ends of the unit cell,
\begin{align}
\label{eqn:x-floquet}
    \mathbf{u}(x_2,y,z)&=\mathbf{u}(x_1,y,z)e^{-\ii k_x (x_2-x_1)}\;,\\
    \mathbf{u}(x,y,z_2)&=\mathbf{u}(x,y,z_1)e^{-\ii k_z (z_2-z_1)}\;,
\end{align}
where $\mathbf{k}=k_x \Hat{\mathbf{e}}_x + k_y \Hat{\mathbf{e}}_y$ is the wave vector. Traction-free boundary conditions are applied on the remaining faces.

\subsection{Embedding the discrete topological models}
\begin{figure*}
    \centering
    \includegraphics[width=\linewidth]{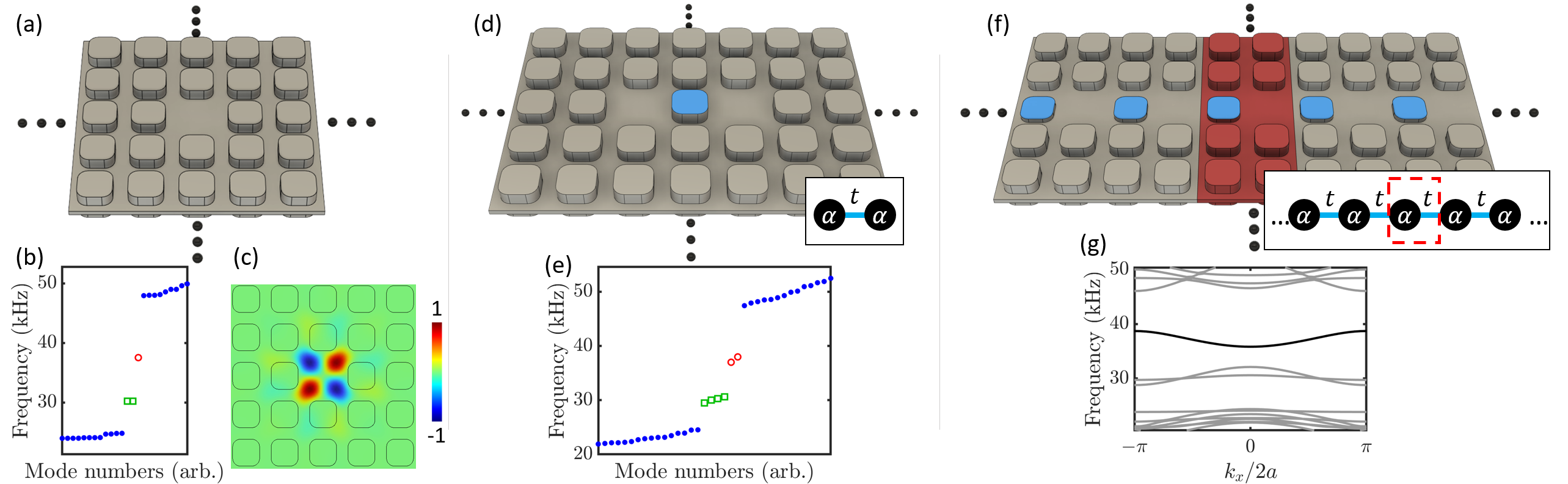}
    \caption{Overview of the design principle based on embedding coupled resonator models into elastic plates. (a) Elastic metamaterial with a point defect in the form of a missing pillar. All metamaterials in this figure have pillars with fillet radius $r=0.25$ in (6.35 mm). (b) Natural frequencies of the metamaterial. Local resonances within the bandgap of the defect-free metamaterial are indicated with open markers. Here, we focus on the local resonance indicated by the red circle. Bulk modes are marked with filled blue circles. (c) Out-of-plane displacement field of the local resonance marked by the red circle in (b). Arbitrary units are used for displacement. (d) Elastic metamaterial with two point defects. The inset shows the effective coupled resonator model corresponding to the dynamically coupled point defects. $\al$ denotes the resonator frequency and $t$ denotes the coupling strength. (e) Natural frequencies of the system. The local resonances of interest are marked with red circles. Other local resonances and bulk modes are marked with green squares and blue dots, respectively. (f) An elastic metamaterial waveguide with point defects arranged in an alternate pattern. The red subdomain represents the unit cell. The inset shows the equivalent coupled resonator model. (g) Dispersion curves of the elastic metamaterial. The black curve arises from the interaction of local resonances. The gray curves are bulk modes or arise from other local resonances; they do not play a role in the design.}
    \label{fig:design-principle-overview}
\end{figure*}

The process of embedding discrete models in the engineered waveguide leverages the concept of coupled local resonances~\cite{escalante_dispersion_2013,wang_collective_2020,jin_physics_2021}, and it is similar to the tight binding method used in quantum mechanics~\cite{simon_oxford_2013}. Since this concept is well-established in the literature, only an overview of its role in the present design is provided (Fig.~\ref{fig:design-principle-overview}). The details of the design, including the effect of the design parameters on the dynamical performance and the role of fillets in the pillars, are discussed in Sec.~II and Sec.~III of the Supplementary Material. 

When one of the pillars of the engineered waveguide is deleted (Fig.~\ref{fig:design-principle-overview}a), the site of the point defect supports a local resonance (Fig.~\ref{fig:design-principle-overview}c) with frequency within the bandgap~\cite{jin_physics_2021} (red circle in Fig.~\ref{fig:design-principle-overview}b). When there are two point defects separated by a pillar (Fig.~\ref{fig:design-principle-overview}d), the local resonances supported by each point defect interact via evanescent coupling~\cite{escalante_dispersion_2013,wang_collective_2020}, which leads to the two modes marked with red circles in Fig.~\ref{fig:design-principle-overview}e. The strength of the coupling is controlled by the height of the intermediate pillar: reducing the pillar height increases the evanescent coupling (Supplementary Material Sec.~II). This system emulates a discrete system of two coupled resonators. Similarly, for a waveguide with a periodic arrangement of point defects separated by a pillar (Fig.~\ref{fig:design-principle-overview}f), the interactions between the local resonances results in a dispersion curve (black curve in Fig.~\ref{fig:design-principle-overview}g), and the system emulates a monoatomic lattice of coupled resonators. In general, by introducing point defects and adjusting pillar heights, a wide variety of discrete systems of resonators with nearest-neighbor couplings can be embedded into the engineered plate

In particular, the design principle can embed periodic arrays of resonators with bipartite coupling schemes into 2D elastic waveguides, which form the basis of BDI class models. Such waveguides preserve time-reversal symmetry and approximately preserve chiral and particle-hole symmetries. Thus, the current design overcomes the challenges outlined at the beginning of the section, and it provides a suitable platform to create a wide array of elastic topological metamaterials of the BDI class. We will apply the design principle to create fully continuous elastic analogs of the SSH and dual SSH models.

\section{Continuous elastic analog of the SSH model}
\label{sec:elastic-ssh}

\begin{figure*}
    \includegraphics[width=\linewidth]{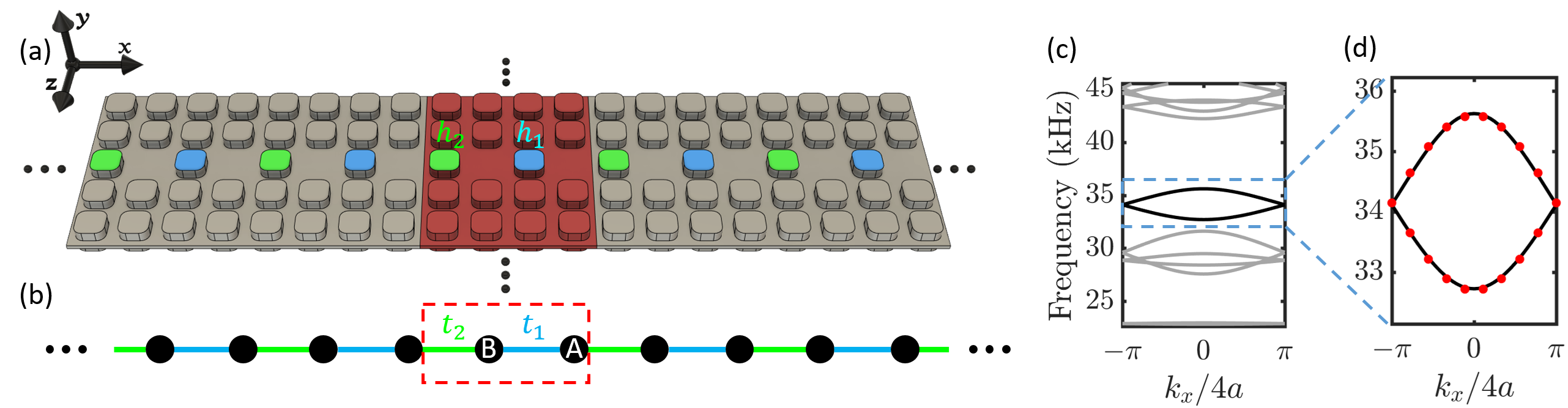}
    \caption{(a) The design of the ESSH(D), where $h_1=h_2=0.3125$ in (7.938 mm). The unit cell used for simulations is marked in red. (b) The schematic of the equivalent SSH chain corresponding to (a). (c) Dispersion relations of the metamaterial. The black curves highlight wave modes resulting from the local resonances. These curves are zoomed into in (d). In (d), the red dots represent the dispersion relation from the coupled resonator approximation.}
    \label{fig:ESSHD}
\end{figure*}

The SSH model consists of a chain of resonators with natural frequency $\al$ that are coupled to their nearest neighbors with alternating strengths $t_1$ and $t_2$ (Sec.~\ref{sec:ssh}). The unit cell consists of two resonators. To create the elastic analogs shown in Figs.~\ref{fig:ESSHD}-\ref{fig:ESSH1}, pillars are deleted in an alternating fashion along a selected row of the waveguide. All pillars have a fillet radius of $r=0.25$ in (6.35 mm). The pillar heights in the selected row alternate between $h_1$ and $h_2$ in order to control the coupling between local resonances and implement the equivalent coupling strengths $t_1$ and $t_2$ of the SSH model.

The values of $h_1$ and $h_2$ are chosen to create topologically distinct elastic analogs by using the phase diagram of the SSH in Fig.~\ref{fig:ssh}b. Identical values of $h_1$ and $h_2$ provide an elastic analog of the SSH model with identical coupling strengths ($t_1=t_2$), which features degeneracies in its dispersion curves. We label this model as ESSH(D), read as \enquote{elastic analog of the SSH with degeneracies.} Then, by perturbing this design to lift the degeneracies, we create elastic analogs of the SSH model with winding numbers $\nu$, called ESSH($\nu$), where $\nu=0,1$.

The computations for all the elastic analogs proceed as follows. The dispersion relation of the proposed design is computed using the finite element method. The unit cell used in the simulation is marked in red in Figs.~\ref{fig:ESSHD}-\ref{fig:ESSH1}. Let its extents in the $xz$ plane be defined by the rectanglar region $[x_1,x_2]\times[z_1,z_2]$. The eigenvalue problem to be solved is defined by the equations of linear elasticity (Eq.~\eqref{eqn:navier-lame}) with periodic boundary conditions on the $xy$ faces at the end of the unit cell,
\begin{equation}
    \mathbf{u}(x,y,z_1)=\mathbf{u}(x,y,z_2)\;,
\end{equation}
Floquet-Bloch boundary conditions on the $yz$ faces at the end of the unit cell (Eq.~\eqref{eqn:x-floquet}), and traction-free boundary conditions on the remaining faces. 

\subsection{ESSH(D)}
Figure~\ref{fig:ESSHD}a presents the design of the ESSH(D) with $h_1=h_2=0.3125$ in (7.938 mm). The unit cell contains two defects. Figure~\ref{fig:ESSHD}c shows the resulting dispersion curves. The two black dispersion curves ($f_{\pm}^{\rm{FEM}}(k_x)$) arise from the local resonances supported by the defects, and they are also plotted separately in Fig.~\ref{fig:ESSHD}d. The curves are nearly symmetric about the frequency $34.12$ kHz, indicating the (approximate) chiral symmetry. The curves are degenerate at $k_x/4a=\pi$ because of the zone-folding effect~\cite{zhang_zone_2019}.

In the range of frequencies of the local resonances (about 32 kHz-36 kHz), the coupled resonator approximation of the ESSH(D) is the SSH model with equal coupling strengths ($t_1=t_2=t$) shown in Fig.~\ref{fig:ESSHD}b. The dispersion curves of the SSH model ($f^{\rm{CR}}_{\pm}(k_x)$) are given by Eq.~\eqref{eqn:ssh-dispersion-relation}, but with $k$ replaced by $k_x/4a$. At $k_x/4a=0$ and $k_x/4a=\pi$,
\begin{equation}
\label{eqn:lsq-fit-ssh-degen}
    \begin{aligned}
        f_{+}^{\rm{CR}}(0) &= \al + 2t\;, \\
        f_{-}^{\rm{CR}}(0) &= \al - 2t\;, \\
        f_{+}^{\rm{CR}}\left( \frac{\pi}{4a} \right) &= \al \;,\\
        f_{-}^{\rm{CR}}\left( \frac{\pi}{4a} \right) &= \al \;. \\
    \end{aligned}
\end{equation}
To find the parameters $\al$ and $t$, we demand $f_{\pm}^{\rm{CR}}(k_x)=f_{\pm}^{\rm{FEM}}(k_x)$ at $k_x/4a=0$ and $\pi$. Since $f_{\pm}^{\rm{FEM}}(k_x)$ is already obtained from the numerical simulation, Eq.~\eqref{eqn:lsq-fit-ssh-degen} defines a system of four linear equations with two unknowns $\al$ and $t$. We compute the least-squares solution of $\al$ and $t$ via MATLAB by using the backslash operator, which results in $\al = 34.15$ kHz and $t = 0.72$ kHz. The resulting dispersion curves $f_{\pm}^{\rm{CR}}(k_x)$ are superimposed on the numerical solutions in Fig.~\ref{fig:ESSHD}d. The agreement between the two sets of curves verifies that the ESSH(D) in Fig.~\ref{fig:ESSHD}a successfully emulates the arrangement of resonators in Fig.~\ref{fig:ESSHD}b. This agreement also implies that the ESSH(D), which is a continuous system, emulates chiral and particle-hole symmetries. 

Nevertheless, since the curves in Fig.~\ref{fig:ESSHD}d do not overlap perfectly, the chiral and particle-hole symmetries are only approximately valid. The imperfect match is attributed to inevitable nonlocal and diagonal evanescent couplings between the point defects. This issue of nonlocal coupling is further discussed in Sec.~III~A of the Supplementary Material.

The ESSH(D) provides the starting point to create the ESSH(0) and ESSH(1). The design strategy is described by the phase diagram of the SSH in Fig.~\ref{fig:ssh}b. The ESSH(D) corresponds to the white circle. Increasing $t_1$ results in the ESSH(0), while increasing $t_2$ results in the ESSH(1). Since increasing $t_i$ implies decreasing $h_i$ in the elastic system, decreasing $h_1$ results in the ESSH(0), while decreasing $h_2$ results in the ESSH(1).

\subsection{ESSH(0)}
\label{sec:essh0}
\begin{figure*}
    \includegraphics[width=\linewidth]{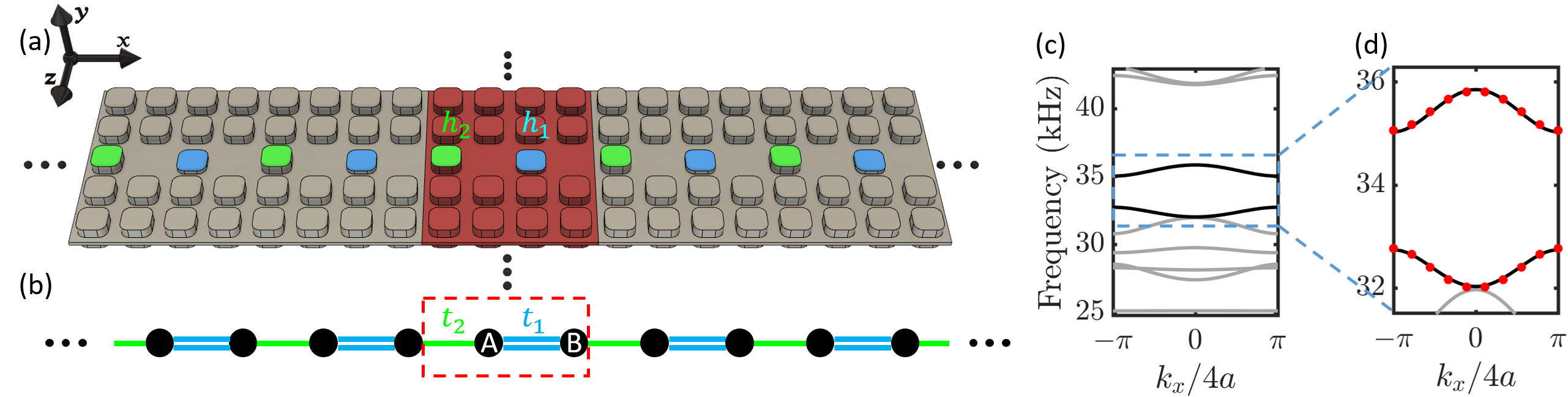}
    \caption{(a) The design of the ESSH($0$), where $h_1=0.125$ in (3.175 mm) and $h_2=0.3125$ in (7.938 mm). The unit cell is marked in red. (b) The schematic of the equivalent SSH chain corresponding to (a). (c) Dispersion relations of the metamaterial. The black curves highlight wave modes resulting from the local resonances. These curves are zoomed into in (d). In (d), the red dots represent the dispersion relation from the coupled resonator approximation.}
    \label{fig:ESSH0}
\end{figure*}

Starting from the ESSH(D), the ESSH(0) is created by decreasing $h_1$. We choose $h_1=0.125$ in (3.175 mm), which corresponds to the smallest value of $h_1$ below which other wave modes strongly interact with the local resonance and prevent the coupled resonator approximation. The resulting design is shown in Fig.~\ref{fig:ESSH0}a. The resulting dispersion curves are shown in Fig.~\ref{fig:ESSH0}c. The dispersion curves $f_{\pm}^{\rm{FEM}}(k_x)$ resulting from the local resonances are shown in Fig.~\ref{fig:ESSH0}d. The two curves are nearly symmetric about a center frequency of $33.89$ kHz and are separated by a bandgap of width $2.29$ kHz. The bandgap width is $6.76\%$ when normalized against the center frequency.

The coupled resonator approximation of the ESSH(0) is the SSH model with nonidentical coupling strengths (Fig.~\ref{fig:ESSH0}b). For the SSH model, the frequencies at $k_x/4a=0$ and $k_x/4a=\pi$ are
\begin{equation}
\label{eqn:lsq-fit-ssh}
    \begin{aligned}
    f_{+}^{\rm{CR}}(0) &= \al + t_1 + t_2\;, \\
    f_{-}^{\rm{CR}}(0) &= \al - t_1 - t_2\;, \\
    f_{+}^{\rm{CR}}\left( \frac{\pi}{4a} \right) &= \al + |t_1 - t_2|\;, \\
    f_{-}^{\rm{CR}}\left( \frac{\pi}{4a} \right) &= \al - |t_1 - t_2|\;,
\end{aligned}
\end{equation}
from Eq.~\eqref{eqn:ssh-dispersion-relation}. In the present design, $h_1 < h_2$, which implies $t_1 > t_2$ and $|t_1 - t_2| = t_1 - t_2$. Consequently, Eqs.~\eqref{eqn:lsq-fit-ssh} are linear in $\al$, $t_1$, and $t_2$. To find the parameters $\al$, $t_1$, and $t_2$, we enforce $f_{\pm}^{\rm{CR}}(k_x)=f_{\pm}^{\rm{FEM}}(k_x)$ at $k_x/4a=0$ and $k_x/4a=\pi$ and find the least squares solution of Eq.~\eqref{eqn:lsq-fit-ssh} using the backslash operator in MATLAB. This provides $\al = 33.91$ kHz, $t_1 = 1.53$ kHz, and $t_2 = 0.38$ kHz. 

The dispersion relation $f_{\pm}^{\rm{CR}}(k_x)$ with these parameters is plotted in Fig.~\ref{fig:ESSH0}d. The agreement between the dispersion curves of the engineered waveguide and its coupled resonator approximation verifies that the ESSH(0) (Fig.~\ref{fig:ESSHD}a) emulates the SSH model (Fig.~\ref{fig:ESSHD}b). Crucially, it follows that the engineered waveguide inherits the winding number of the corresponding SSH model, which equals 0 because $t_1>t_2$ (Eq.~\eqref{eqn:ssh-winding-number}). Thus, the ESSH(0) is indeed a continuous system with winding number 0.

\subsection{ESSH(1)}
\begin{figure*}
    \includegraphics[width=\linewidth]{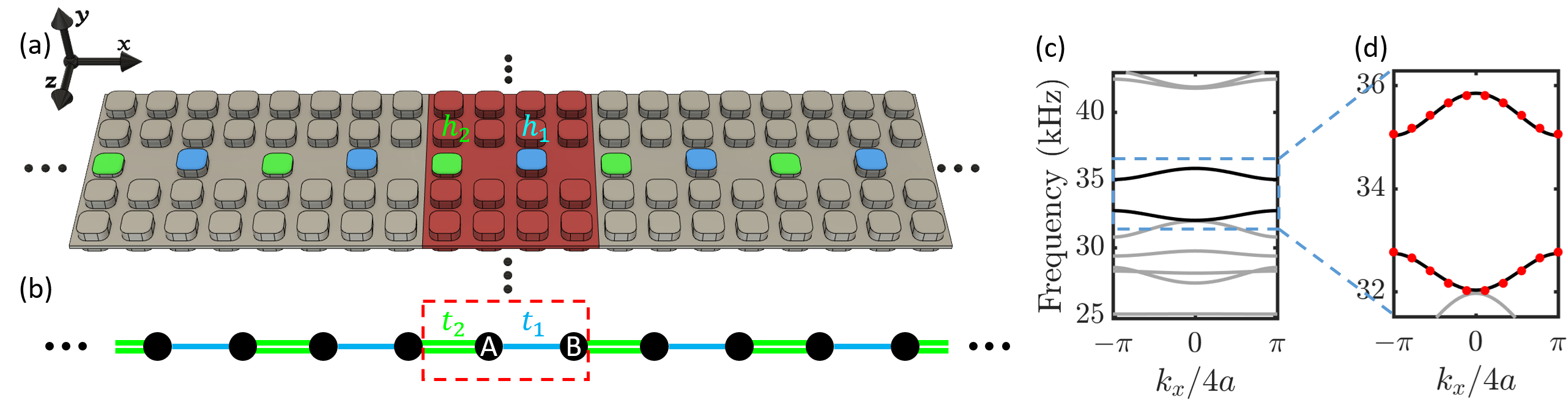}
    \caption{(a) The design of the ESSH($1$), where $h_1=0.3125$ in (7.938 mm) and $h_2=0.125$ in (3.175 mm). The unit cell is marked in red. (b) The schematic of the equivalent SSH chain corresponding to (a). (c) Dispersion relations of the metamaterial. The black curves highlight wave modes resulting from the local resonances. These curves are zoomed into in (d). In (d), the red dots represent the dispersion relation from the coupled resonator approximation.}
    \label{fig:ESSH1}
\end{figure*}

The ESSH(1) is created from the ESSH(D) by decreasing $h_2$ to $0.125$ in (3.175 mm). The resulting design is shown in Fig.~\ref{fig:ESSH1}a and its dispersion curves are shown in Figs.~\ref{fig:ESSH1}c,d. By the same fitting technique used for the ESSH(0), but noting that $h_1 > h_2$ in the present design, we find $\al = 33.91$ kHz, $t_1 = 0.38$ kHz, and $t_2 = 1.53$ kHz. Since $t_1 < t_2$, the winding number of the SSH model equals $1$ by Eq.~\eqref{eqn:ssh-winding-number}, implying that the winding number of the engineered waveguide also equals $1$. 

Notice that the design of the ESSH(1) is identical to ESSH(0) except for the choice of the unit cell. As a result, the dispersion curves of the ESSH(1) shown in Figs.~\ref{fig:ESSH1}c,d are identical to Figs.~\ref{fig:ESSH0}c,d. However, the guided mode displacement profiles are not identical (they differ by a translation), which leads to the different topological properties~\cite{yin_band_2018}. 

\subsection{Edge modes}

Finite realizations of the ESSH can support topological edge modes, in analogy to its discrete counterpart (Sec.~\ref{sec:ssh}). The bulk-boundary correspondence predicts that topological edge modes arise at the end of a finite ESSH(1) design and at an interface between the ESSH(0) and ESSH(1) designs~\cite{chiu_classification_2016}. The emergence of edge modes in these systems is numerically verified in this section by computing natural frequencies and mode shapes. The simulations are performed using the finite element method on the commercial software COMSOL Multiphysics 6.1.     

\subsubsection{Truncated ESSH(1)}
\label{sec:essh-edge-mode-free}
A discrete SSH model with winding number 1 supports an edge mode at its free end (Sec.~\ref{sec:ssh}). A free end in the resonator model corresponds to \enquote{empty space,} but in the context of the present waveguides, it means the absence of defects. Thus, a discrete SSH model with a free end corresponds to a finite ESSH that continues into a 2D waveguide without defects, instead of the usual free boundary condition of an elastic waveguide.

\begin{figure*}
    \includegraphics[width=\linewidth]{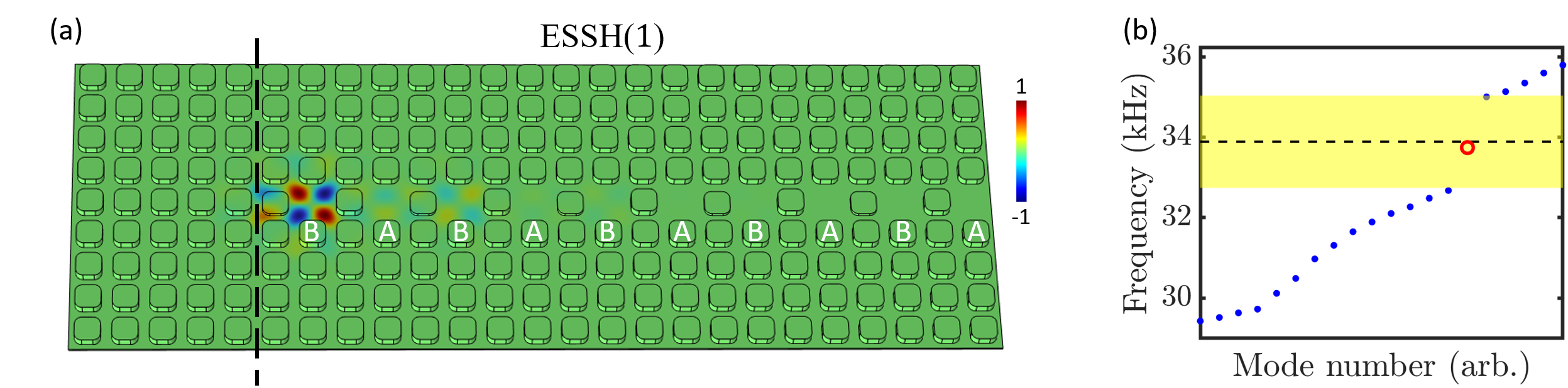}
    \caption{(a) A metamaterial waveguide created by joining finite segments of the ESSH(1) and the defect-free metamaterial. The interface is marked by the dashed line. The out-of-plane displacement field of the topological edge mode is shown by the color map. Arbitrary displacement units are used. The letter adjacent to a point defect marks its sublattice. (b) The natural frequencies of the metamaterial around $\al=33.91$ kHz. The red circle and blue dots represent the topological edge mode and bulk modes. The yellow rectangle highlights the bandgap of the ESSH(1). The dashed black line marks $\al=33.91$ kHz, which is the predicted frequency of the topological edge mode.}
    \label{fig:ESSH-edge-mode-free}
\end{figure*}

With this knowledge, the edge mode at the end of the ESSH(1) can be realized in the metamaterial in Fig.~\ref{fig:ESSH-edge-mode-free}a. The metamaterial is created from a lattice with 9 rows and 25 columns of pillars. Its shorter edges are free and longer edges are fixed. The first 5 columns are free of defects and emulate the free end condition for resonators. In the 6th to 25th columns, pillars are deleted and their heights are adjusted to create five unit cells of the ESSH($1$).

The natural frequencies around $\al=33.91$ kHz are plotted in Fig.~\ref{fig:ESSH-edge-mode-free}b. There is one mode with natural frequency 33.74 kHz that lies within the bandgap of the infinite ESSH(1) (32.74 kHz-35.04 kHz). This is the topological edge mode predicted by the bulk-boundary correspondence~\cite{chiu_classification_2016}. The proximity between the natural frequency of the edge mode, 33.74 kHz, and $\al=33.91$ kHz is a signature of the chiral and particle-hole symmetries approximately preserved by the waveguide. 

The displacement field of the topological edge mode plotted in Fig.~\ref{fig:ESSH-edge-mode-free}a also shows features of the chiral and particle-hole symmetries. Recall that the local resonance at a point defect has a characteristic antisymmetric shape with two perpendicular nodal lines separating four antinodes (Fig.~\ref{fig:design-principle-overview}c). As shown in Sec.~IV of the Supplementary Material, the displacement amplitudes of these local resonances at the point defects can be considered as the discrete degrees of freedom of the continuous system, analogous to the tight-binding approximation in quantum mechanics~\cite{simon_oxford_2013}. Then, the displacement amplitudes are (to first order) nonzero for sublattice B and zero for sublattice A. From this approximate perspective, the chiral and particle-hole operators (from Eq.~\eqref{eqn:chiral-operator}) reverse the degrees of freedom of sublattice A, leaving the topological edge mode invariant.

\subsubsection{Interface between the ESSH(0) and the ESSH(1)} 
\begin{figure*}
    \includegraphics[width=\linewidth]{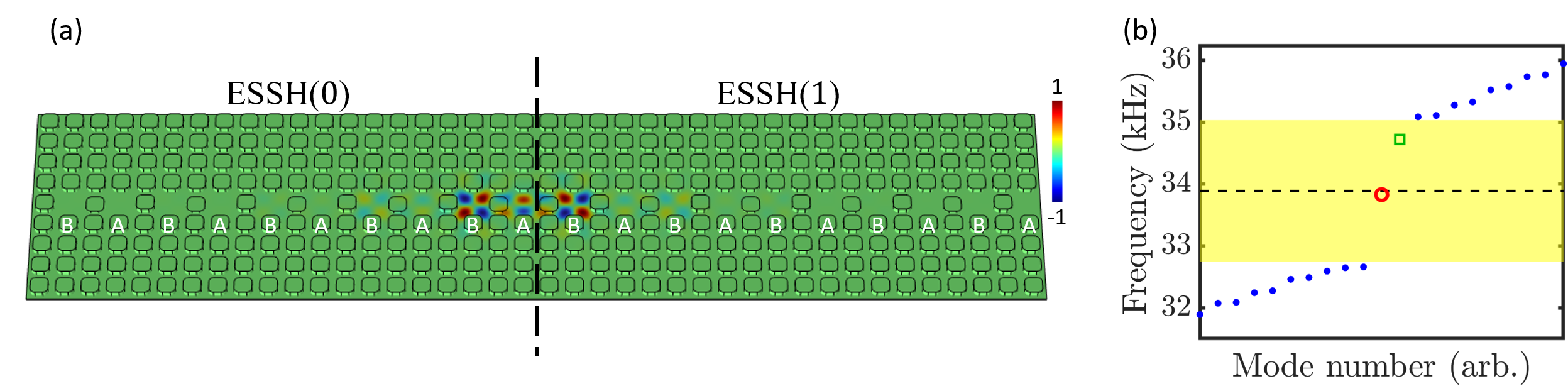}
    \caption{(a) A metamaterial waveguide created by joining finite segments of the ESSH($0$) and ESSH($1$). The interface is marked by the dashed line. The out-of-plane displacement field of the topological edge mode is shown by the color map. Arbitrary units of displacement are used. The letter adjacent to a point defect marks its sublattice. (b) The natural frequencies of the metamaterial centered about $\al=33.91$ kHz. The red circle, green square, and blue dots represent the topological edge mode, an edge mode at the right end, and bulk modes. The yellow rectangle highlights the common bandgap supported by the ESSH(0) and ESSH(1). The dashed black line marks $\al=33.91$ kHz, which is the predicted frequency of the topological edge mode.}
    \label{fig:ESSH-edge-mode-10}
\end{figure*}

An interface between the ESSH(0) and the ESSH(1) is realized using the metamaterial shown in  Fig.~\ref{fig:ESSH-edge-mode-10}a. The metamaterial is created from a lattice with 9 rows and 40 columns of pillars. All its edges are fixed. In the left half of the fifth row, pillars are deleted and their heights are adjusted to create five unit cells of the ESSH($0$). Similarly, the right half of the fifth row is tailored to emulate five unit cells of the ESSH($1$). Thus, the central row supports an interface between the ESSH($0$) and ESSH($1$). 

The natural frequencies of the metamaterial centered around $\al=33.91$ kHz are plotted in Fig.~\ref{fig:ESSH-edge-mode-10}b. There are two modes with natural frequencies within the frequency range 32.74 kHz-35.04 kHz, which is the bandgap of the infinite elastic analog of the SSH. Only one of these modes is localized at the interface of the two SSH chains (Fig.~\ref{fig:ESSH-edge-mode-10}a), as predicted by the bulk-boundary correspondence. The other mode is localized at the left boundary. The latter mode does not arise from topological considerations because the fixed boundary condition of the waveguide does not translate into a free boundary condition of the equivalent resonator model. 

The presence of approximate chiral and particle-hole symmetries influences the frequency and mode shape of the topological edge mode. The frequency of the topological edge mode, 33.83 kHz, is close to the center frequency $\al=33.91$ kHz. The displacement field of the topological edge mode is plotted in Fig.~\ref{fig:ESSH-edge-mode-10}a. In the first-order approximation, the mode shape consists of local resonances at sublattice B. Thus, the chiral and particle-hole operators approximately leave the mode shape invariant.

\section{Elastic analog of the dual SSH model}
\label{sec:elastic-kitaev}
As previously discussed, the dual SSH model can be implemented by leveraging two identical but staggered SSH chains coupled to each other (Sec.~\ref{sec:dual-ssh}). To create the elastic analogs in Figs.~\ref{fig:EDSSH0}-\ref{fig:EDSSHn1}, we first select two rows separated by a row of pillars. We delete alternate pillars along these two rows, so that each row emulates an SSH system. The remaining pillars in the two rows have alternating heights, $h_1$ (blue pillar) and $h_2$ (green pillar), which control the intra-SSH couplings $t_1$ and $t_2$ (Sec.~\ref{sec:elastic-ssh}). The alternating pillar heights follows a staggered scheme between the two rows: if the pillar heights in the first row are $[h_1,h_2,\dots]$, the corresponding pillar heights in the second row are $[h_2,h_1,\dots]$. The (orange) pillars between adjacent defects of different rows have height $h_c$. They control the inter-SSH coupling $t_c$. The pillars that mediate the coupling, that is, those with heights $h_1$, $h_2$, and $h_c$, have a fillet radius $r'$. The remaining pillars have a fillet radius $r$. 

To create topologically distinct elastic analogs, we use the phase diagram of the dual SSH model (Fig.~\ref{fig:dssh-plots}b) to choose suitable values of the pillar heights. First, we choose $h_1=h_2=h_c$ so that $t_1=t_2$ by the symmetry of the system and $t_c$ is approximately equal to $t_1$ and $t_2$.  This design is denoted as the EDSSH(D), read as \enquote{elastic analog of the dual SSH with degeneracies.} Then, the design is perturbed to lift the degeneracies and obtain elastic analogs of the dual SSH model with winding numbers $\nu=0,1,-1$, denoted EDSSH($\nu$). The designs for EDSSH($\nu$) must support bandgaps centered at identical frequencies for the bulk-boundary correspondence to hold. We ensure this condition by choosing the fillet radii appropriately. 

The finite element simulations of the unit cell and of the finite system, as well as the least squares approach (to fit the unknown parameters of the coupled resonator approximation) that will be discussed in the following paragraphs use the same procedures outlined in Sec.~\ref{sec:elastic-ssh} and, for brevity, will not be discussed again.

\subsection{EDSSH(D)}
\begin{figure*}
    \centering
    \includegraphics[width=\linewidth]{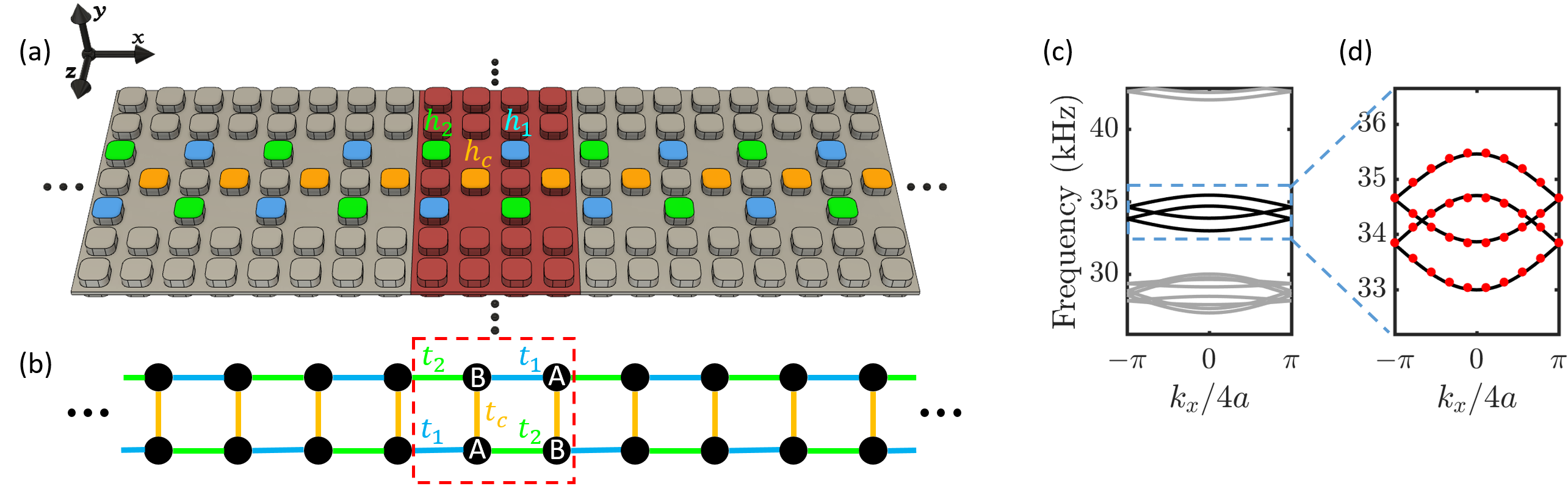}
    \caption{(a) The design of the EDSSH(D), where $h_1=h_2=h_c=0.3125$ in (7.938 mm). The unit cell used for simulations is marked in red. (b) The schematic of the equivalent dual SSH model corresponding to (a). (c) Dispersion relations of the metamaterial. The black curves highlight wave modes resulting from the local resonances. These curves are zoomed into in (d). In (d), the red dots represent the dispersion relation from the coupled resonator approximation.}
    \label{fig:EDSSHD}
\end{figure*}

Figure~\ref{fig:EDSSHD}a shows the design for the EDSSH(D), where $h_1=h_2=h_c=0.3125$ in (7.938 mm) and $r=r'=0.25$ in (6.35 mm). One unit cell contains four defects. Figure~\ref{fig:EDSSHD}c shows the dispersion relation of the metamaterial. The four black dispersion curves arise from the local resonances supported by the defects, and they are plotted separately in Figure~\ref{fig:EDSSHD}d. The dispersion curves are nearly symmetric about $34.26$ kHz, which is a feature of the (approximate) chiral symmetry. The dispersion curves are two-fold degenerate at $k_x/4a=\pm\pi$ and at $k_x/4a \approx \pm 2.14$. The degeneracy at $k_x/4a=\pm \pi$ is because of the zone-folding effect~\cite{zhang_zone_2019} (Sec.~\ref{sec:dual-ssh}). The curves cross at $k_x/4a \approx \pm 2.14$ because the system is symmetric under a reflection about a plane parallel to the $xy$ plane and containing the center of the unit cell. 

The coupled resonator approximation of the continuous system is the dual SSH model with equal intra-SSH coupling strengths ($t_1=t_2=t$) shown in Fig.~\ref{fig:EDSSHD}b. The dispersion curves of the discrete system are given by Eq.~\eqref{eqn:dssh-dispersion-relation}, but with $k$ replaced by $k_x/4a$. At $k_x/4a=0$,
\begin{equation}
\label{eqn:lsq-fit-EDSSHd}
    \begin{aligned}
    f_1^{\rm{CR}}(0)&=\al - t_c - 2t\;, \\
    f_2^{\rm{CR}}(0)&=\al + t_c - 2t\;, \\
    f_3^{\rm{CR}}(0)&=\al - t_c + 2t\;, \\
    f_4^{\rm{CR}}(0)&=\al + t_c + 2t\;,
\end{aligned}
\end{equation}
in the order of ascending natural frequencies. The least-squares fit of the natural frequency of the resonators $\al$, intra-SSH coupling strength $t$, and inter-SSH coupling strength $t_c$ provides $\al = 34.26$ kHz, $t=0.41$ kHz, and $t_c=0.41$ kHz. The dispersion curves obtained from the coupled resonator approximation are superimposed over the numerical solutions in Fig.~\ref{fig:EDSSHD}d. The agreement between the two sets of curves verifies that the EDSSH(D) in Fig.~\ref{fig:EDSSHD}a emulates the system of coupled resonators in Fig.~\ref{fig:EDSSHD}b.

The EDSSH(D) serves as the foundation to create designs for the EDSSH($0$), EDSSH($1$), and EDSSH($-1$). The design strategy follows from the topological phase diagram of the dual SSH model in Fig.~\ref{fig:dssh-plots}b. The EDSSH(D) corresponds to the white circle. Starting from the EDSSH(D), the EDSSH($0$), EDSSH($1$), and EDSSH($-1$) can be created by increasing $t_c$, $t_2$, and $t_1$, respectively. Recall that to increase the coupling strength, the corresponding pillar height is decreased.

\subsection{EDSSH($0$)}
\begin{figure*}
    \centering
    \includegraphics[width=\linewidth]{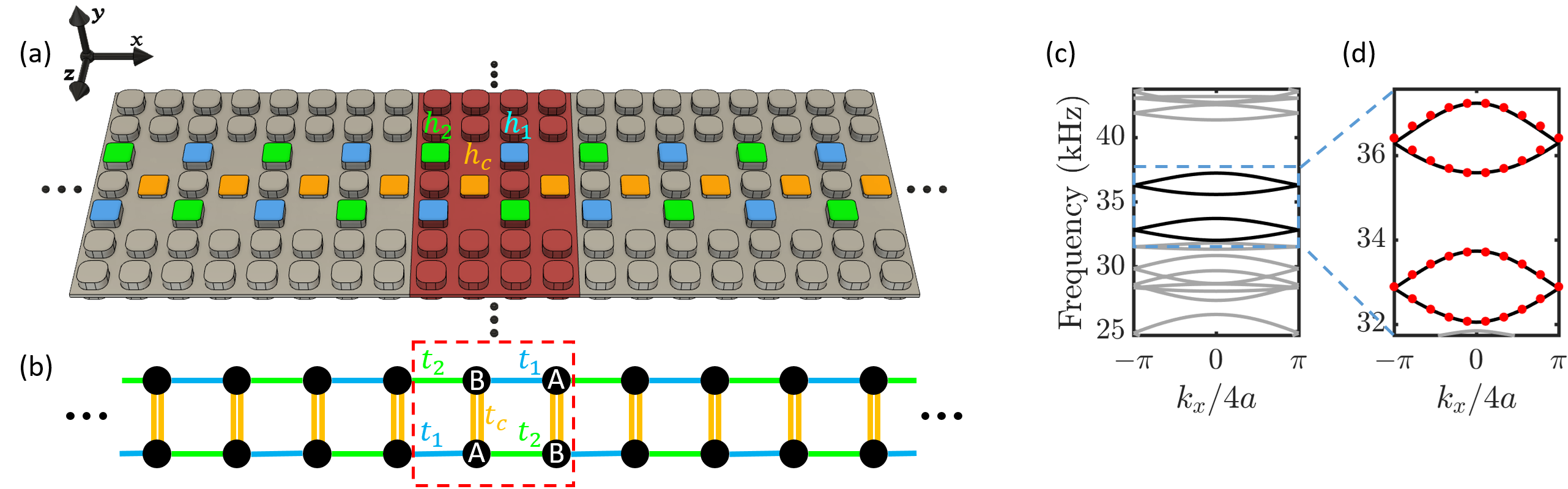}
    \caption{(a) The design of the EDSSH($0$), where $h_1=0.3125$ in (7.938 mm), $h_2=0.3125$ in (7.938 mm), and $h_c=0.125$ in (3.175 mm). The unit cell is marked in red. (b) The schematic of the equivalent dual SSH model corresponding to (a). (c) Dispersion relations of the metamaterial. The black curves highlight wave modes resulting from the local resonances. These curves are zoomed into in (d). In (d), the red dots represent the dispersion relation from the coupled resonator approximation.}
    \label{fig:EDSSH0}
\end{figure*}

To create the EDSSH($0$) from the EDSSH(D), $h_c$ is decreased until a bandgap opens in the dispersion curves of the local resonances, leading to the design in Fig.~\ref{fig:EDSSH0}a. The pillar heights are $h_1=h_2=0.3125$ in (7.938 mm) and $h_c=0.125$ in (3.175 mm). The numerical value of $h_c$ was chosen to create the largest bandgap possible before other wave modes started interacting with the local resonances. We choose the fillet radii as $r=0.3$ in (7.62 mm) and $r'=0.125$ in (3.175 mm).

Figure~\ref{fig:EDSSH0}c shows the dispersion curves of the EDSSH($0$). Figure~\ref{fig:EDSSH0}d separately plots the dispersion curves resulting from the local resonances. There is a bandgap centered at $34.66$ kHz. The width of the bandgap is $1.86$ kHz; its normalized width is $5.36$ \%.

The coupled resonator approximation of the EDSSH(0) is the dual SSH model with identical intra-SSH couplings strengths ($t_1=t_2=t$) shown in Fig.~\ref{fig:EDSSH0}b. By using the same fitting process used for the EDSSH(D), we find $\al=34.66$ kHz, $t= 0.42$ kHz, and $t_c = 1.76$ kHz.

The dispersion curves of the coupled resonator approximation is superimposed on the numerically obtained curves in Fig.~\ref{fig:EDSSH0}d. The agreement between the two sets of dispersion curves verifies that the EDSSH(0) in Fig.~\ref{fig:EDSSH0}a emulates the discrete dual SSH in Fig.~\ref{fig:EDSSH0}b, implying that the engineered waveguide inherits the winding number of the discrete system. The discrete dual SSH with parameters $t_1=t_2=0.42$ kHz and $t_c = 1.76$ kHz has a winding number of 0 by Eq.~\eqref{eqn:dual-ssh-winding-number}. Thus, the EDSSH(0) also has a winding number of 0.

\subsection{EDSSH(1)}
\begin{figure*}
    \centering
    \includegraphics[width=\linewidth]{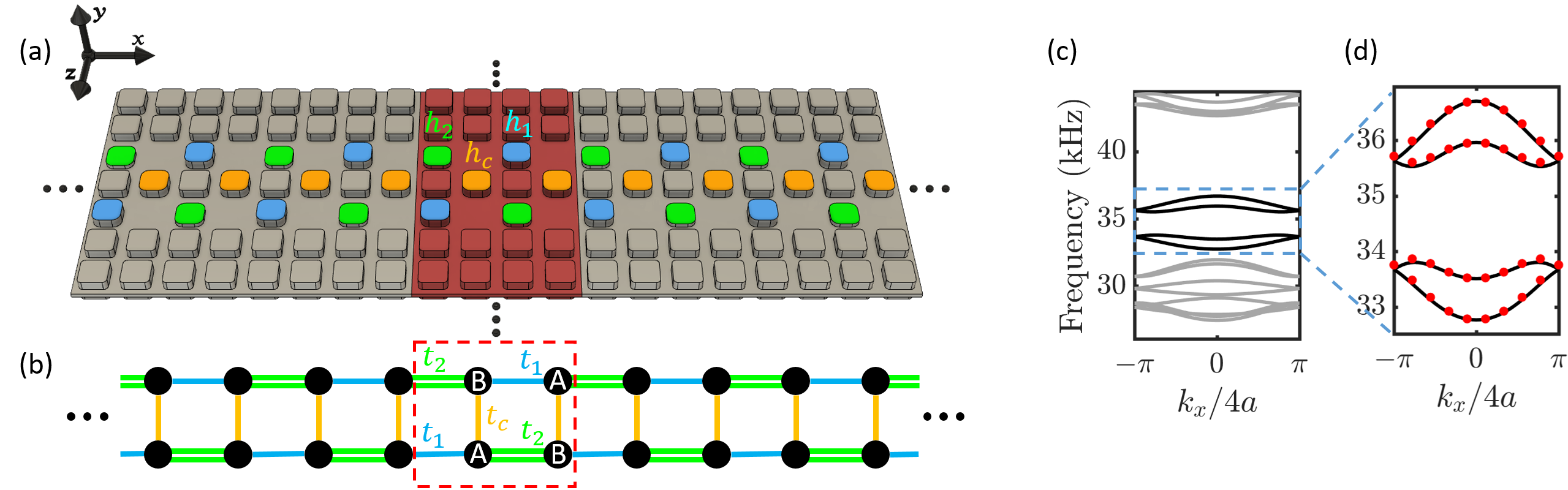}
    \caption{(a) The design of the EDSSH($1$), where $h_1=0.3125$ in (7.938 mm), $h_2=0.1406$ in (3.572 mm), and $h_c=0.3125$ in (7.938 mm). The unit cell is marked in red. (b) The schematic of the equivalent dual SSH model corresponding to (a). (c) Dispersion relations of the metamaterial. The black curves highlight wave modes resulting from the local resonances. These curves are zoomed into in (d). In (d), the red dots represent the dispersion relation from the coupled resonator approximation.}
    \label{fig:EDSSH1}
\end{figure*}

The EDSSH($1$) is created from the EDSSH(D) by decreasing $h_2$. The design in Fig.~\ref{fig:EDSSH1}a uses $h_1=0.3125$ in (7.938 mm), $h_2=0.1406$ in (3.572 mm), and $h_c=0.3125$ in (7.938 mm). The fillet radii are $r=0.125$ in (3.175 mm) and $r'=0.25$ in (6.35 mm). 

Figure~\ref{fig:EDSSH1}c shows the dispersion curves for the EDSSH($1$). The dispersion curves arising from the local resonances are plotted separately in Fig.~\ref{fig:EDSSH1}d, where the second and third dispersion curves are separated by a bandgap centered at $34.67$ kHz. The bandgap is $1.72$ kHz wide; its normalized width is $4.98 \%$.

The coupled resonator approximation of the EDSSH($1$) design is the dual SSH model shown in Fig.~\ref{fig:EDSSH1}b. According to Eq.~\eqref{eqn:dssh-dispersion-relation}, the frequencies of the wave modes in ascending order at $k_x/4a=0$ are 
\begin{equation}
\label{eqn:lsq-fit-EDSSH0}
    \begin{aligned}
        f_1^{\rm{CR}}(0)&=\al - t_c - (t_1+t_2)\;, \\
        f_2^{\rm{CR}}(0)&=\al + t_c - (t_1+t_2)\;, \\
        f_3^{\rm{CR}}(0)&=\al - t_c + (t_1+t_2)\;, \\
        f_4^{\rm{CR}}(0)&=\al + t_c + (t_1+t_2)\;.
    \end{aligned}
\end{equation}
A least squares fit between Eqs.~\eqref{eqn:lsq-fit-EDSSH0} and the numerical solutions provides $\al = 34.74$ kHz, $t_c = 0.37$ kHz, and $(t_1+t_2) = 1.60$ kHz.

$(t_1-t_2)$ still needs to be found. From the frequencies of the wave modes at $k_x/4a=\pi$,
\begin{equation}
\label{eqn:fit-EDSSH-dt}
        |t_1 - t_2| = \sqrt{|(f_i^{\rm{CR}}(\pi)-\al)^2 - t_c^2|}
\end{equation}
for $i=1,\dots,4$. Assuming $f_i^{\rm{CR}}(\pi/4a)$ equals the numerical values at $k_x/4a=\pi$, we take the average of the four values from Eq.~\eqref{eqn:fit-EDSSH-dt} and find that $|t_1-t_2|=0.91$ kHz. The relative magnitudes of $t_1$ and $t_2$ are known from the heights $h_1$ and $h_2$: here $h_1 > h_2$, so $t_1 < t_2$. Thus, we find $(t_1-t_2) = -|t_1 - t_2| = -0.91$ kHz. This provides $t_1=0.35$ kHz and $t_2=1.25$ kHz. 

The dispersion relation from the coupled resonator approximation is plotted in Fig.~\ref{fig:EDSSH1}d, which agrees well with the numerically obtained curves. Thus, the EDSSH(1) in Fig.~\ref{fig:EDSSH1}a emulates the dual SSH model in Fig.~\ref{fig:EDSSH1}b. Since the winding number of the discrete dual SSH model equals $1$ by Eq.~\eqref{eqn:dual-ssh-winding-number}, the winding number of the EDSSH(1) also equals $1$.

\subsection{EDSSH($-1$)}
\begin{figure*}
    \centering
    \includegraphics[width=\linewidth]{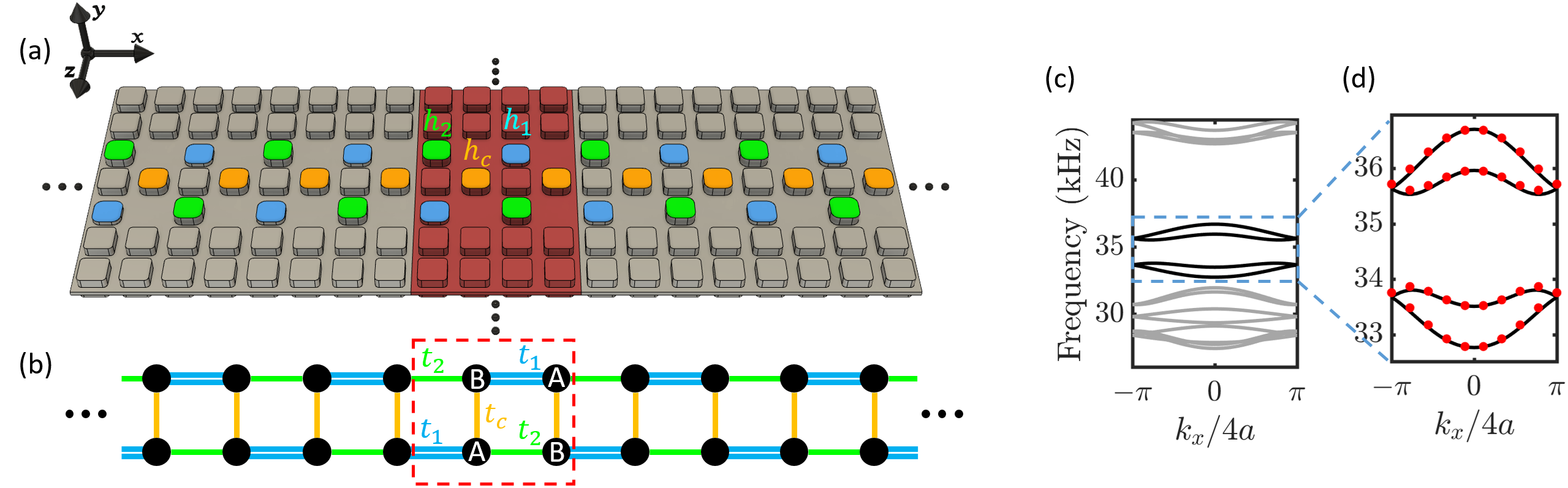}
    \caption{(a) The design of the EDSSH($-1$), where $h_1=0.1406$ in (3.572 mm), $h_2=0.3125$ in (7.938 mm), and $h_c=0.3125$ in (7.938 mm). The unit cell is marked in red. (b) The schematic of the equivalent dual SSH model corresponding to (a). (c) Dispersion relations of the metamaterial. The black curves highlight wave modes resulting from the local resonances. These curves are zoomed into in (d). In (d), the red dots represent the dispersion relation from the coupled resonator approximation.}
    \label{fig:EDSSHn1}
\end{figure*}

The EDSSH($-1$) is created from the EDSSH(D) by decreasing $h_1$. The design in Fig.~\ref{fig:EDSSHn1}a uses $h_1=0.1406$ in (3.572 mm), $h_2=0.3125$ in (7.938 mm), and $h_c=0.3125$ in (7.938 mm). The fillet radii are $r=0.125$ in (3.175 mm) and $r'=0.25$ in (6.35 mm). 

The dispersion curves of the metamaterial are shown in Figs.~\ref{fig:EDSSHn1}c,d. They are identical to the dispersion curves for the EDSSH($1$) because the two metamaterials differ only by a choice of unit cell. By the same fitting strategy used for the EDSSH($1$), but noting that $h_1 < h_2$, we find $\al = 34.74$ kHz, $t_c = 0.37$ kHz, $t_1=1.25$ kHz and $t_2=0.35$ kHz. As expected, the values of $t_1$ and $t_2$ are exchanged when compared with the EDSSH($1$). The dispersion relations from the coupled resonator approximation are plotted in Fig.~\ref{fig:EDSSHn1}d, which agree with the numerically obtained curves. Since the discrete dual SSH model with $t_1=1.25$ kHz, $t_2=0.35$ kHz, and $t_c = 0.37$ kHz has a winding number of $-1$, the winding number of the EDSSH($-1$) also equals $-1$.

\subsection{Edge modes}
Finite realizations of the EDSSH can support topological edge modes in accordance with the bulk-boundary correspondence~\cite{chiu_classification_2016}. Such edge modes are supported at five configurations: (i) at an end of the EDSSH(1), (ii) at an end of the EDSSH($-1$), (iii) at an interface between EDSSH(0) and EDSSH(1), (iv) at an interface between EDSSH(0) and EDSSH($-1$), and (v) at an interface between EDSSH(1) and EDSSH($-1$). Since the designs for the EDSSH(1) and EDSSH($-1$) differ by a reflection alone (about a plane parallel to the $xy$ plane) and the design for the EDSSH(0) is invariant under this reflection, configurations (i) and (ii) and configurations (iii) and (iv) display very similar properties. Thus, we only investigate the edge modes in configurations (i), (iii), and (v).

\subsubsection{Truncated EDSSH(1)}
\label{sec:edssh-free-edge}
\begin{figure*}
    \includegraphics[width=\linewidth]{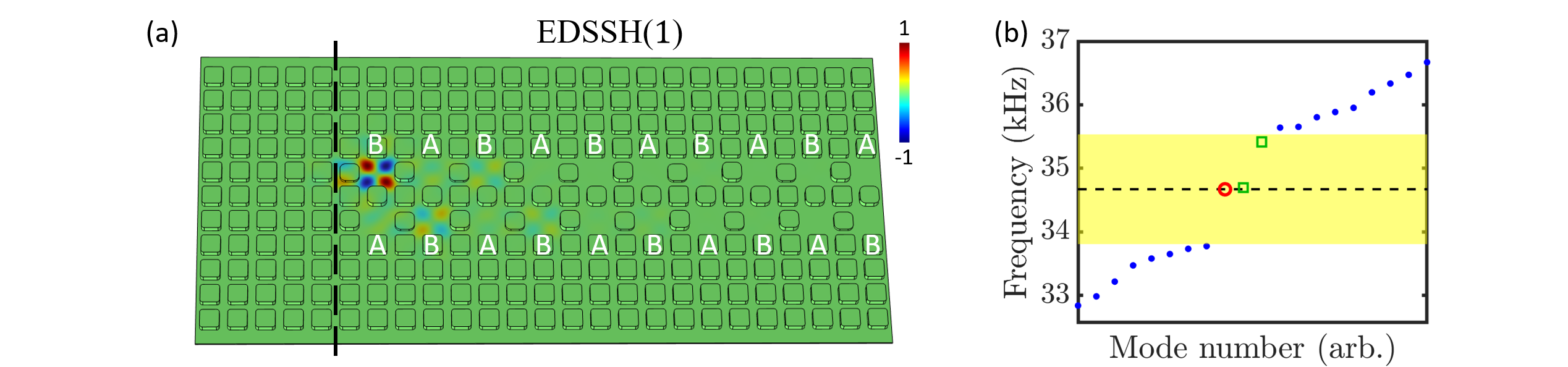}
    \caption{(a) A metamaterial waveguide created by joining finite segments of the EDSSH(1) and the defect-free metamaterial. The interface is marked by the dashed line. The out-of-plane displacement field of the topological edge mode is shown by the color map. Arbitrary units of displacement are used. The letter adjacent to a point defect marks its corresponding sublattice. (b) The natural frequencies of the metamaterial centered about $\al=34.74$ kHz. The red circle, green squares, and blue dots represent the topological edge mode, edge modes localized at the right end, and bulk modes. The yellow rectangle highlights the bandgap of the ESSH(1). The dashed black line marks $\al=34.74$ kHz, which is the predicted frequency of the topological edge mode.}
    \label{fig:EDSSH-edge-mode-free}
\end{figure*}

Noting that the free end in a discrete resonator model corresponds to the EDSSH continuing into the engineered waveguide without defects, the edge mode at the end of the EDSSH(1) is realized using the metamaterial in Fig.~\ref{fig:EDSSH-edge-mode-free}a. The metamaterial is created from a lattice with 11 rows and 25 columns of pillars. Pillars in the first 5 columns of the emulate the free boundary condition of a resonator system. Pillars in the 6th to 25th columns of the three central rows are selectively deleted and their heights are adjusted to create five unit cells of the EDSSH(1).The longer edges of the metamaterial are fixed; the shorter edges are free.

The natural frequencies of the system centered around $\al=34.74$ kHz are plotted in Fig.~\ref{fig:EDSSH-edge-mode-free}b. There are three modes in the frequency range 33.81 kHz-35.53 kHz, which is the bandgap in the design of the EDSSH(1) of infinite extent. Only one of these modes with frequency 34.69 kHz (marked with a red circle) is localized at the left end of the EDSSH(1). This is the topological edge mode as predicted by the bulk-boundary correspondence, whose mode shape is shown in Fig.~\ref{fig:EDSSH-edge-mode-free}a. The other two modes (marked with green squares) are localized at the free end. They do not have a topological origin because the free boundary condition of the waveguide does not correspond to a free boundary condition in the resonator model. 

The topological edge mode exhibits characteristics resulting from the (approximate) chiral and particle-hole symmetries of the continuous system. The frequency of the mode, 34.69 kHz, is close to $\al=34.74$ kHz. The mode shape is primarily comprised of local resonances only in sublattice B. The nonzero displacements at point defects of sublattice A result from mode leakage and imperfect chiral symmetry in the continuous system, as discussed in Sec.~\ref{sec:essh-edge-mode-free}. 

The equivalence between the dual SSH model and the Kitaev chain model (Sec.~\ref{sec:dual-ssh} and Supplementary Material Sec.~I) implies that the topological edge mode of the EDSSH(1) is a Majorana-like mode supported at the free end of a classical emulation of a Kitaev chain. If the local resonances at each point defect are viewed as the degrees of freedom, the topological edge mode of the EDSSH(1) maps to the topological edge mode of the dual SSH, which in turn maps to the Majorana-like mode of the Kitaev chain (Supplementary Material Sec.~IV~A). With this identification, the invariance of the Majorana-like mode of the Kitaev chain under the particle-hole operator can also be observed in the topological edge mode of the EDSSH(1). The particle-hole operator in the EDSSH(1) reverses the response of the local resonance degrees of freedom of sublattice B. Since the response is localized on sublattice B, it is invariant under the particle-hole operator.

\subsubsection{Interface between the EDSSH(0) and the EDSSH(1)}
\begin{figure*}
    \includegraphics[width=\linewidth]{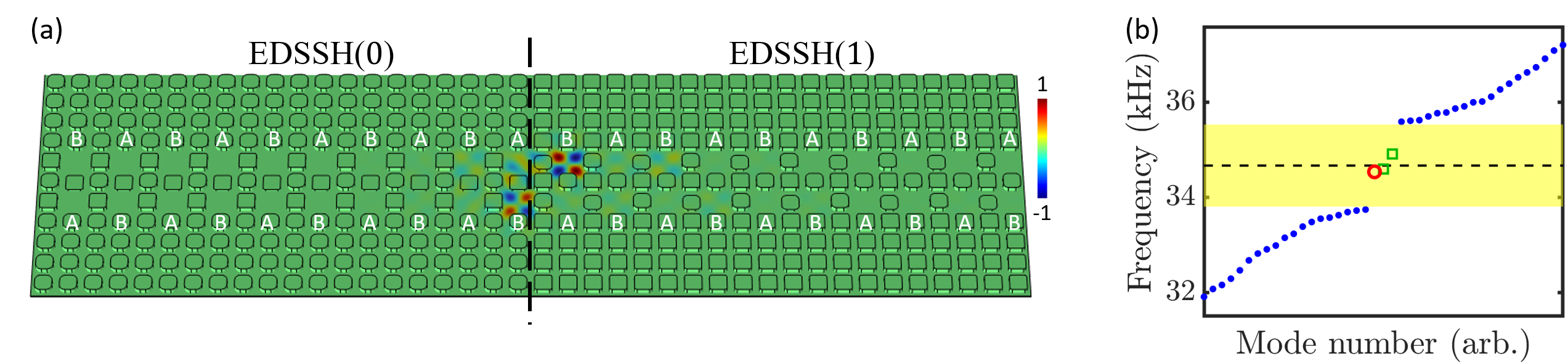}
    \caption{(a) A metamaterial created by joining finite segments of the EDSSH($0$) and EDSSH($1$) exhibiting a topological edge mode. The EDSSH($0$)-EDSSH($1$) interface is marked by the dashed line. The color map indicates the out-of-plane displacement field in arbitrary units. The letter adjacent to a point defect marks its corresponding sublattice. (b) The natural frequencies of the metamaterial centered about $34.70$ kHz. The red circle, green squares, and blue dots represent the topological edge mode, edge modes at the right end, and bulk modes. The yellow rectangle highlights the common bandgap supported by the EDSSH($0$) and EDSSH($1$). The dashed black line marks $34.70$ kHz, which is the expected frequency of the topological edge mode.}
    \label{fig:EDSSH-edge-mode-10}
\end{figure*}

An interface between the EDSSH(0) and the EDSSH(1) is created in the metamaterial shown in Fig.~\ref{fig:EDSSH-edge-mode-10}a. The metamaterial is created from a plate with 11 rows and 40 columns of pillars. Pillars in the three central rows are selectively deleted and their heights are adjusted to create five unit cells of the EDSSH(0) design followed by five unit cells of the EDSSH(1) design. All ends of the metamaterial are fixed.

The natural frequencies of the system are shown in Fig.~\ref{fig:EDSSH-edge-mode-10}b. There are three modes in the frequency range $33.81$ kHz-$35.53$ kHz, which is the common bandgap between the two designs. However, only the mode marked with a red circle is a topological edge mode localized at the interface of interest. The other two modes are localized at the fixed end. These modes do not have a topological origin as the fixed boundary condition of the elastic waveguide does not correspond to a free or fixed boundary condition in the equivalent resonator model.

The topological edge mode is influenced by the (approximate) chiral and particle-hole symmetries of the continuous system. Its frequency is $34.54$ kHz, which is close to the average values of $\al$ of the EDSSH(0) and EDSSH(1), $34.70$ kHz. To a first-order approximation, its mode shape (Fig.~\ref{fig:EDSSH-edge-mode-10}a) is a superposition of local resonances in sublattice B (Supplementary Material Sec.~IV~B). Thus, it is invariant under the chiral and particle-hole operations. By the same argument of Sec.~\ref{sec:edssh-free-edge}, the topological edge mode is a Majorana-like mode, which maps to the Majorana-like mode supported at the interface of classical analogs of a topologically trivial and a topologically nontrivial Kitaev chain.

\subsubsection{EDSSH(1)-EDSSH($-$1) interface}
\begin{figure*}
    \includegraphics[width=\linewidth]{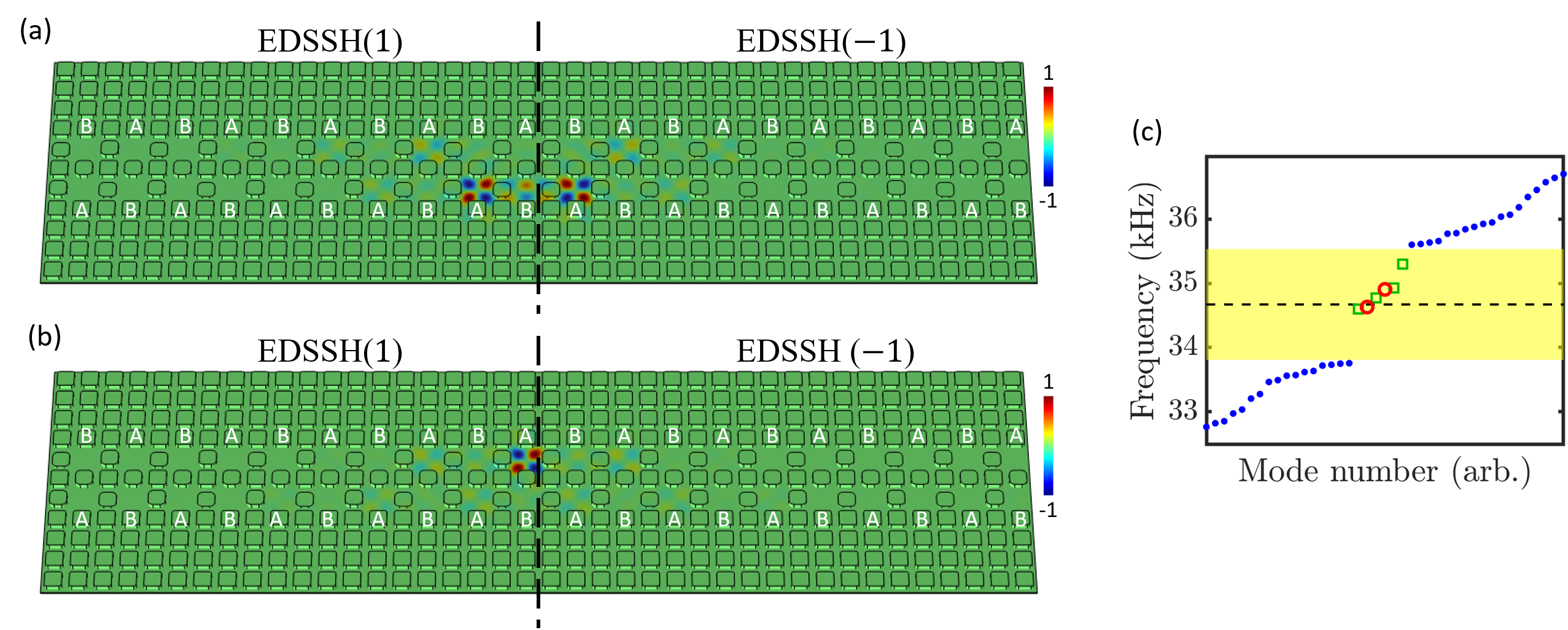}
    \caption{(a,b) A metamaterial created by joining finite segments of the EDSSH($1$) and EDSSH($-1$) exhibiting two topological edge modes. The EDSSH($1$)-EDSSH($-1$) interface is marked by the dashed line. The color map indicates the out-of-plane displacement field in arbitrary units. The letter adjacent to a point defect marks its corresponding sublattice. (c) The natural frequencies of the metamaterial centered about $34.74$ kHz. The red circle, green squares, and blue dots represent the topological edge mode, edge modes at the left or right ends, and bulk modes. The yellow rectangle highlights the common bandgap supported by the EDSSH($-1$) and EDSSH($1$). The dashed black line marks $34.74$ kHz, which is the predicted frequency of the topological edge mode.}
    \label{fig:EDSSH-edge-modes-1n1}
\end{figure*}

An interface between the EDSSH(1) and the EDSSH($-1$) is created in the metamaterial shown in Fig.~\ref{fig:EDSSH-edge-modes-1n1}a. The metamaterial is created from a plate with 11 rows and 40 columns of pillars. Pillars in the three central rows are selectively deleted and their heights are adjusted to create five unit cells of the EDSSH(1) design followed by five unit cells of the EDSSH($-1$) design. All ends of the metamaterial are fixed.

The results for this interface are shown in Fig.~\ref{fig:EDSSH-edge-modes-1n1}. Figure~\ref{fig:EDSSH-edge-modes-1n1}c shows the natural frequencies. There are six modes in the common bandgap of the two designs, $33.81$ kHz-$35.53$ kHz. Only two of these are topological edge modes at the central interface, in accordance with the bulk-boundary correspondence. The other modes are localized at the left or right ends and do not have a clear topological interpretation.

The topological edge modes are influenced by the (approximate) chiral and particle-hole symmetries of the system. Their frequencies are $34.62$ kHz and $34.91$ kHz, which is close to $\al=34.74$ kHz. Their mode shapes are plotted in Fig.~\ref{fig:EDSSH-edge-modes-1n1}a and Fig.~\ref{fig:EDSSH-edge-modes-1n1}b, which, to first order, are a superposition of local resonances in sublattice A. Thus, the mode shapes are approximately invariant under chiral and particle-hole operations.

\section{Experimental validation}
\label{sec:experiments}
\begin{figure*}
    \centering
    \includegraphics[width=\linewidth]{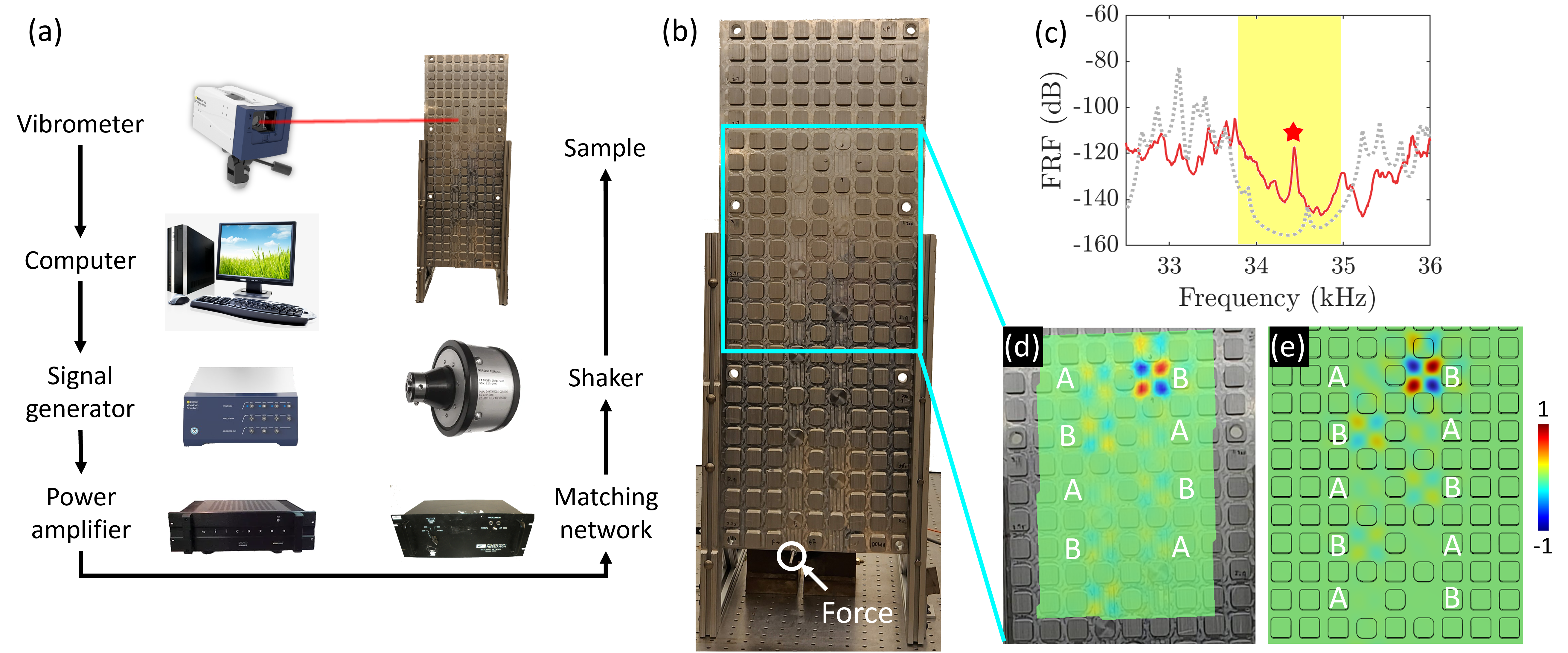}
    \caption{(a) Schematic illustration of the different components of the experimental setup. (b) Test sample of the truncated EDSSH(1) subjected to a point excitation (see label \enquote{Force}). (c) Comparison of the experimental (solid red line) and numerical (dotted gray line) frequency response function. The yellow rectangle indicates the topological bandgap observed in the experimental frequency response function. The star denotes the frequency of occurrence of the topological edge mode. (d) Measured out-of-plane displacement field (operating deflection shape) associated with the topological edge mode. (e) Numerically simulated out-of-plane displacement field associated with the topological edge mode. The plot is reproduced from Fig.~\ref{fig:EDSSH-edge-mode-free}a. The displacements in (d) and (e) are scaled so that the largest magnitude is 1.}   
    \label{fig:experimental-validation}
\end{figure*}

The presence of a Majorana-like mode was experimentally verified for the case of the truncated EDSSH(1) (Sec.~\ref{sec:edssh-free-edge}). The experimental setup is illustrated in Fig.~\ref{fig:experimental-validation}a. The EDSSH(1) was fabricated from a plate of aluminum 6061 using a Computer Numerical Control (CNC) milling machine. The final sample (Fig.~\ref{fig:experimental-validation}b) was 12 in (304.80 mm) wide and 31.25 in (793.75 mm) long. A section of length 19.09 in (485 mm) on the longer edges of the sample was clamped via a frame to an optical table. This length was arbitrarily chosen based on the size of the available frame; however, this length negligibly influences the dynamics of the topological mode because the mode is localized far from the boundaries of the sample. A point on the lower edge of the sample located at approximately 4.75 in (120.65 mm) from the left edge (labeled \enquote{Force} in Fig.~\ref{fig:experimental-validation}b) was connected to a piezoelectric shaker (Wilcoxon F7) via a threaded rod. The shaker applied a driving force to the metamaterial plate in the out-of-plane direction. The input to the shaker was generated by the internal signal generator of a scanning laser vibrometer Polytec PSV 500. The signal was then amplified via a Wilcoxon PA8HF amplifer and fed into the shaker via a matching network (Wilcoxon N7FS). The response was measured using the scanning vibrometer.

The frequency response function (FRF) of the metamaterial waveguide was determined in response to a white noise excitation over a range 30 kHz - 40 kHz and voltage amplitude 0.5 V (at the output of the signal generator). The displacement frequency spectrum was normalized against the displacement at the driving point, and ensemble averaging was performed over the acquisition points contained within the first four rows of defects. The FRF is plotted in Fig.~\ref{fig:experimental-validation}c over the range 32.5kHz - 36kHz, where the dynamics of the plate is dictated by local resonances. The FRF indicates the presence of a bandgap that consists of a single topological edge mode, in agreement with the modal analysis performed in Fig.~\ref{fig:EDSSH-edge-mode-free}d. The experimental FRF (red line) qualitatively agrees with the numerically computed FRF (dashed gray line). The numerical FRF was obtained by performing a frequency domain analysis using COMSOL Multiphysics. The material parameters used in the numerical analysis were density of 2700 kg/m$^3$, Young's modulus of elasticity of 69 GPa, Poisson's ratio 0.33, and an isotropic loss factor of 0.0007 to better match the experimental results. Further, the plate thickness $h_\mathrm{plate}$ was set to 0.126 in (3.2 mm) and the plate thickness of the last unit cell was reduced to 0.098 in (2.5 mm) to model manufacturing flaws in the actual experimental sample. This flaw perturbs the chiral symmetry of the system and reduces the width of the bandgap; nevertheless, the topological mode of the system persists~\cite{shi_disorder-induced_2021}, therefore highlighting the typical robustness of topological dynamics.

The operating deflection shape associated with the topological edge mode is found experimentally by exciting the structure periodically at the corresponding natural frequency (39.43 kHz, red star in Fig.~\ref{fig:experimental-validation}c) and using a voltage amplitude of 0.25 V at the signal generator. The measured out-of-plane displacement field is plotted in Fig.~\ref{fig:experimental-validation}d, and it agrees well with the result of the numerical simulation shown in Fig.~\ref{fig:experimental-validation}e. A pronounced localized displacement profile is confined to sublattice B, in strong agreement with the discrete dual SSH and the Majorana-like nature of the mode. 

\section{Conclusions}
\label{sec:conclusions}
This study developed a concept for the realization of continuous elastic topological metamaterials based on design principles rooted in resonantly coupled waveguides. The design approach allows leveraging existing 1D discrete topological models (such as the SSH chain and ladder) and embed them into continuous waveguides. The proposed design uses an intuitive and scalable pillared plate configuration to create the foundational gapped metamaterial within which the 1D-inspired topological models are embedded. Point and line defects are leveraged to create local resonators and quasi-1D waveguides, while the pillars' height provides a simple and effective way to tune the coupling strength between them. This design approach allows implementing classical analogs of chiral and particle-hole symmetries, which ultimately leads to continuous elastic topological metamaterials belonging to the BDI class. It is highlighted that, while the pillared design was selected in this study, the approach presented is general and can be combined with most design strategies.

More specifically, the design methodology was used to obtain elastic topological metamaterial plates capable of emulating the SSH chain and ladder. Their topological properties were verified analytically via the winding number invariant as well as numerically and experimentally by observing the emergence of edge modes at topologically non-trivial interfaces. The edge modes displayed an almost exact midgap frequency and their associated eigenstates were localized to a single sublattice; both behaviors are in full agreement with the occurrence of chiral and particle-hole symmetries. Furthermore, edge modes occurring in the elastic analog of the dual SSH model were shown to be Majorana-like modes. 

In conclusion, the general design strategy presented in this study provides a systematic method to embed a wide array of topological properties in continuous elastic systems. This approach suggest pathways towards exciting structural applications including, but not limited to, localized vibrations in systems with large disorder~\cite{shi_disorder-induced_2021} and fault-tolerant on-material computation using Majorana-like modes~\cite{barlas_topological_2020,chenTopologicalComputationNonAbelian2025}.

\begin{acknowledgments}
The authors gratefully acknowledge the partial financial support of the National Science Foundation under grant \#2330957.
\end{acknowledgments}

\bibliography{References}

\end{document}